\documentclass{article}
\usepackage{arxiv}

\usepackage[utf8]{inputenc} 
\usepackage[T1]{fontenc}    
\usepackage{hyperref}       
\usepackage{url}            
\usepackage{booktabs}       
\usepackage{amsfonts}       
\usepackage{nicefrac}       
\usepackage{microtype}      
\usepackage{lipsum}
\usepackage{graphicx}
\usepackage{upgreek}
\graphicspath{ {./images/} }

\usepackage{cite}
\usepackage{amsmath,amssymb,amsfonts}
\usepackage{algorithmic}
\usepackage{graphicx}
\usepackage{textcomp}
\usepackage{chemmacros}
\usepackage{array}
\usepackage{subcaption}
\usepackage{caption}
\usepackage{balance}

\title{Understanding the Security Landscape of Embedded Non-Volatile Memories: A Comprehensive Survey}

\author{
Zakia Tamanna Tisha \\
  Department of Electrical and Computer Engineering\\
  Auburn University\\
  Auburn, AL 36849\\
  \texttt{zakia.tisha@auburn.edu} \\
   \And
 Ujjwal Guin \\
  Department of Electrical and Computer Engineering\\
  Auburn University\\
  Auburn, AL 36849 \\
  \texttt{ujjwal.guin@auburn.edu} \\
  }
       
\begin{document}
\maketitle

\begin{abstract}
The modern semiconductor industry requires memory solutions that can keep pace with the high-speed demands of high-performance computing. Embedded non-volatile memories (eNVMs) address these requirements by offering faster access to stored data at an improved computational throughput and efficiency. Furthermore, these technologies offer numerous appealing features, including limited area-energy-runtime budget and data retention capabilities. Among these, the data retention feature of eNVMs has garnered particular interest within the semiconductor community. Although this property allows eNVMs to retain data even in the absence of a continuous power supply, it also introduces some vulnerabilities, prompting security concerns. These concerns have sparked increased interest in examining the broader security implications associated with eNVM technologies. This paper examines the security aspects of eNVMs by discussing the reasons for vulnerabilities in specific memories from an architectural point of view. Additionally, this paper extensively reviews eNVM-based security primitives, such as physically unclonable functions and true random number generators, as well as techniques like logic obfuscation. The paper also explores a broad spectrum of security threats to eNVMs, including physical attacks such as side-channel attacks, fault injection, and probing, as well as logical threats like information leakage, denial-of-service, and thermal attacks. Finally, the paper presents a study of publication trends in the eNVM domain since the early 2000s, reflecting the rising momentum and research activity in this field.

\end{abstract}

\keywords{Non-volatile memories\and security primitives\and PUFs\and TRNGs\and side-channel analysis\and probing\and fault injection}

\section{Introduction}
The demand for memory devices has evolved significantly over the past few decades. This shift is primarily driven by factors such as scaling and the demand for high-density and faster memory devices. Traditional memory technologies like SRAM and DRAM have been the building blocks of computing systems for years. Although they provide fast memory access, their biggest limitation lies in their dependence on continuous power to retain information. Flash memory revolutionized storage by enabling mobility applications. However, as semiconductor scaling reaches its physical limits, flash memory struggles to meet the market demand for more endurance, write speed, and power efficiency. These limitations have set the stage for next-generation eNVM technologies, offering significant advantages such as non-volatility and enhanced performance. With their advanced architectures, eNVMs are becoming essential to modern high-performance computing systems. 

Beyond performance enhancements, eNVMs also serve as integral components in secure hardware systems. They facilitate the design of secure architectures that resist tampering and are used in a variety of applications, including cryptography and secure boot processes. They support scalability, reconfigurability, and cost-effective local computing. They also serve as a rich source of entropy, making them ideal candidates for building security primitives like physically unclonable functions (PUFs) and true random number generators (TRNGs). Nevertheless, the integration of eNVMs into modern computing systems also introduces new risks, such as the potential leakage of sensitive data like secret keys, login credentials, and credit card information~\cite{khan2021comprehensive}. Historically, data security in volatile memories (like SRAMs and DRAMs) was not a major concern since they lose their content when powered down. Yet, as these cache memories evolve to potentially become non-volatile, there is an increased risk of adversaries accessing sensitive information in its unencrypted form. As the memory landscape continues to evolve, a comprehensive understanding of both aspects of eNVM security has become crucial to developing reliable computing environments.

Devices used for security purposes increasingly rely on the unique properties of eNVMs. Their non-volatility enables critical data to persist across power cycles, ensuring secure operation even during power outages. eNVMs also offer tamper resistance for protecting sensitive data like cryptographic keys from physical and logical attacks. Additionally, eNVMs have increased endurance to support frequent updates and record security data, and their fast access speeds help retrieve important information quickly. They also include advanced error correction to maintain data integrity and prevent corruption.

In this paper, we dive deep into the security aspects of eNVMs. Our study begins with an in-depth look at various eNVM technologies, exploring the underlying technology behind security devices and the solutions that leverage eNVMs. As the applications of eNVMs grow, so do the risks associated with them. A key part of our research also involves exploring the structural characteristics of eNVMs that make them susceptible to security attacks. Compared to the previous work published in ISVLSI~\cite{tisha2024exploring}, we make the following modifications in this paper:

\begin{itemize}

   \item \textit{Expanding Vulnerability Study:}  This study builds on our preliminary research, which explored five different non-volatile memory (NVM) technologies. Here, we provide a more detailed analysis of their security vulnerabilities. In particular, we investigate the architectural configurations and critical parametric choices that make eNVMs susceptible to various security attacks. For example, factors such as array topology, write current, material properties, etc., are shown to impact eNVM security significantly.
    
   \item \textit{Broadening Security Primitive Study:} In this work, we explore the technologies underlying security devices and solutions that leverage eNVMs. Building upon the extensive literature reviewed in the preliminary study, additional publications are examined to advance the collective understanding of the field further. The findings indicate that the unique physical and electrical properties of eNVMs enable the development of various security primitives, including PUFs, TRNGs, and logic obfuscation techniques.

   \item \textit{Expanding Attack Vectors:}  We investigate additional security attacks targeting eNVMs that were not addressed in our initial work. These newly examined attacks include information leakage attacks, which exploit side channels or remanence to expose sensitive data; denial of service (DoS) attacks, which aim to disrupt memory operations and impact system availability;  and thermal attacks, where temperature manipulation is used to trigger faults or change memory behavior. Examining these security threats provides a broader perspective of the security challenges associated with eNVM-based systems.
   
   \item \textit{Analysis of Research Trends:} 
   This study presents a comprehensive analysis of publication trends in the NVM domain since the early 2000s, highlighting the evolution of research focus over time. There has been a significant surge in research activity across diverse NVM technologies, driven by rising interest and accelerated progress in the field. In parallel, we examine the key technological milestones that illustrate how NVM technologies have evolved over the years to address challenges such as energy efficiency, performance scalability, and computing efficiency. By capturing both research trends and innovation trajectories, the study offers valuable insights into the evolution of NVM technologies, helping to shape future directions and innovations in memory design, particularly in the security domain.

\end{itemize}

The rest of the paper is structured as follows: Section~\ref{sec:NVM-types} covers various NVM technologies and discusses their architectural vulnerabilities. Section~\ref{sec:NVM-security} delves into security applications that utilize eNVMs. Section~\ref{sec:NVM-attacks} examines a broad spectrum of attack vectors targeting eNVMs. Section~\ref{sec:nvm-trends} presents a detailed analysis of NVM research trends. Section~\ref{sec:NVM-future} considers potential future developments in this field. The paper concludes with Section~\ref{sec:conclusion}.

\section{Background}\label{sec:NVM-types} 
Traditionally, the combination of CMOS-based memories, such as volatile DRAM and SRAM, and non-volatile flash has been adequate to meet both the temporary and permanent data storage requirements of multi-chip systems. The trend towards system-on-chip integration with scalability, reconfigurability, and very low power has driven the development of additional eNVM technologies with new memory and computational architectures. These memory types utilize specific materials that have the ability to maintain a bistable state in their electronic characteristics. This distinguishing property enables data retention in eNVMs without requiring continuous electrical power for several years. Until recently, there have been minimal changes to the fundamental technology and cells responsible for retaining data across power cycles. Currently, floating gate or oxide-nitride-oxide trapped charge (ONO) cell structures are the predominant core technologies in the majority of eNVM devices~\cite{derhacobian2010power}. 

eNVMs are ideal for a wide range of applications, from consumer electronics to high-performance computing and embedded systems. In addition to the familiar characteristics of speed, density, and power consumption, retention and endurance are two crucial metrics for evaluating NVMs. Retention is the ability of a memory cell to retain its contents over a period of time. Endurance is the number of write and erase operations that can be performed before the quality of the cell degrades as a result of wear-out. They are important because wear-out is usually stronger in NVMs than their volatile counterparts and depends on the duty cycle and security pattern~\cite{gerardin2010present}.

Depending on the underlying technology, NVM exhibits unique structural attributes that influence its performance and susceptibility to specific fault mechanisms. This section provides an overview of the architectural foundations and intrinsic vulnerabilities of five widely studied NVMs - flash memory, phase change memory (PCM), magnetoresistive random access memory (MRAM), resistive random access memory (RRAM), and ferroelectric random access memory (FeRAM). The discussion highlights their operational principles, material compositions, and structural limitations that pose security challenges.

\begin{figure}[t]
\centering
\includegraphics[width=0.85\linewidth]{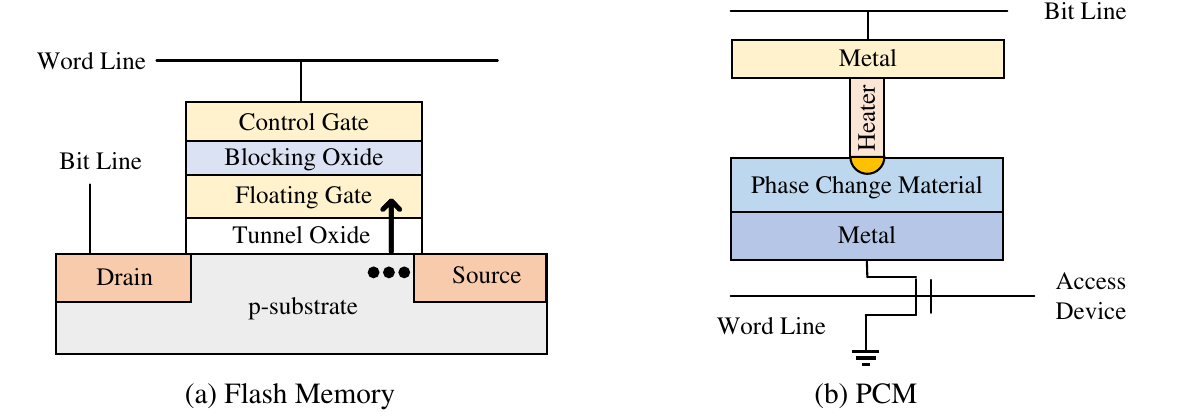} \vspace{-5px}
\caption{Bitcell diagram of a (a) Flash Memory~\cite{vatajelu2014nonvolatile} and (b) PCM~\cite{aswathy2021future}.}   \label{fig:nvm-fig-1}
\vspace{-10px}
\end{figure}

\subsection{Flash Memory} A non-volatile device based on Metal-Oxide-Semiconductor Field-Effect Transistor (MOSFET) technology~\cite{bez2003introduction} and shown in Figure~\ref{fig:nvm-fig-1}(a), Flash memory has revolutionized electronic devices. It utilizes floating gate memory for information storage and tunneling current for programming and erasing. Charge injection or removal from the floating gate enables state retention even after power removal. Flash memory is extensively used in medical diagnostic systems, digital cameras, and mobile phones due to its non-volatility, magnetic immunity, and compatibility with current CMOS processes. However, scaling may face limitations due to tunnel oxide constraints and the cost of integration.

The flash memory architecture has a thin tunnel oxide that supports efficient carrier transport. This thin oxide layer is susceptible to reliability issues like reduced operation voltage and deterioration after numerous program and erase cycles. Researchers have been exploring alternative technologies like nitride-based memory, nanocrystal memory, etc., as promising candidates~\cite{meena2014overview}.

\subsection{Phase Change Memory}  PCM, also known as PCRAM, is a type of non-volatile RAM characterized by a simple capacitor-like structure (shown in Figure~\ref{fig:nvm-fig-1}(b)), with a thin chalcogenide semiconductor film sandwiched between electrodes, facilitating easy miniaturization~\cite{fujisaki2012overview}. These devices boast long cycle life, low programming energy, and excellent scaling characteristics. Chalcogenide phase-change materials, commonly containing elements from group 6 of the periodic table and further expandable to additional material systems by doping, are prominent in PCM, with $GeSbTe$ alloys, especially the GST pseudobinary composition, showing high promise. Operating on the principle of phase change from amorphous to crystalline or vice versa, PCM undergoes this transition at a relatively low temperature of around 600\textdegree{C}, driven by energy from Joule heat generated by current passing through the PCM cell. The resistivity of chalcogenide material varies between the crystal and amorphous phases, allowing data storage based on resistivity changes.

PCM suffers from limited write endurance, with cells enduring only about $10^7$ to $10^9$ writes, making them prone to wear-out attacks. Additionally, resistance drift in Multi-level Cell (MLC) PCM can cause transient errors over time, further compromising data integrity. Furthermore,  PCM is vulnerable to tampering attacks, such as magnetic or thermal manipulation, which can alter memory content or prolong data retention for unauthorized access~\cite{wang2019threat}.

\subsection{Magnetoresistive Random Access Memory}  
\begin{figure}[t]
\centering
\includegraphics[width=0.9\linewidth]{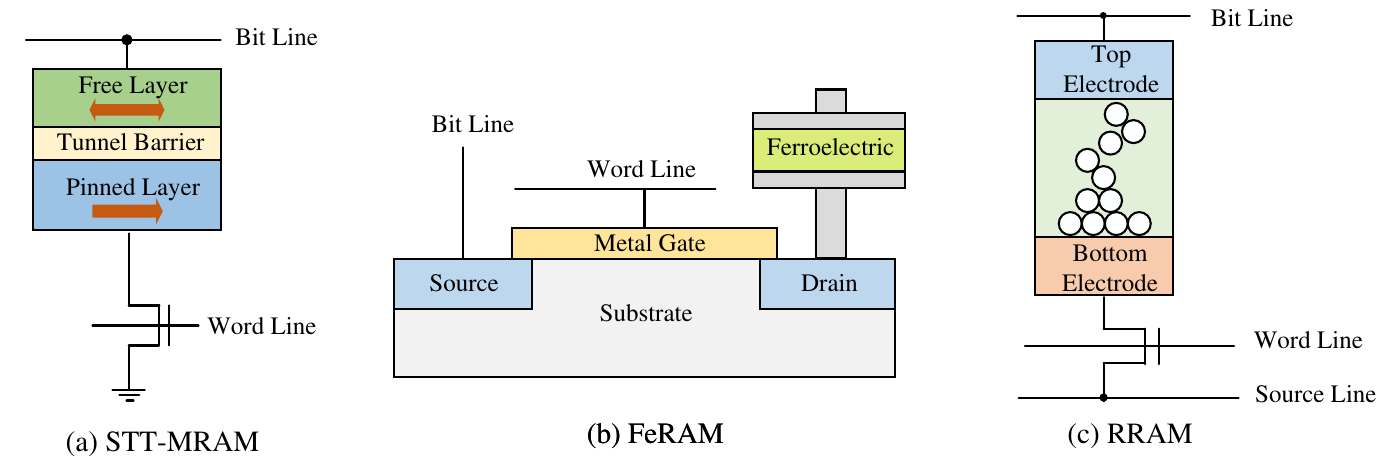} \vspace{-5px}
\caption{Bitcell diagram of a (a) STT-MRAM~\cite{khan2021study}, (b) RRAM~\cite{khan2021study}, (c) FeRAM~\cite{meena2014overview}. 
}   \label{fig:nvm-fig-2}
\vspace{-10px}
\end{figure}

MRAM has been prevalent since the 90s. It is a type of non-volatile memory ideal for high-density applications like solid-state disks. The MRAM architecture is a unique combination of spintronic devices with silicon-based microelectronics. It contains two magnetic storage elements stacked on each other and separated by a thin insulating tunnel barrier- these magnetic plates and the insulating layer form the magnetic tunnel junction (MTJ). One of the magnetic plates forms the fixed layer, whose magnetic direction always stays the same. The other plate forms the free layer, whose magnetic direction changes according to the bias applied to the MTJ~\cite{meena2014overview}. A magnetoresistance effect called tunnelling magnetoresistance (TMR) occurs in the MTJ. The thin insulating layer allows electrons to tunnel through it from one plate to the other. When the magnetic layers are parallel, the cell has a low resistance. On the contrary, the cell is in a high resistance state if they are antiparallel. The resistance state determines whether the binary bit stored in the MRAM is a 1 or a 0. MRAM can be further categorized into spin-transfer torque MRAM (STT-MRAM) and spin-orbit torque MRAM (SOT-MRAM) based on the torque mechanisms employed for switching. STT-MRAM has seen widespread research and commercialization, whereas SOT-MRAM is emerging as a promising successor, offering potential improvements in switching speed and endurance. STT-MRAM addresses high operating current issues by manipulating magnetization direction in the free layer using spin-polarized current between layers. This technology promises low-current, cost-effective MRAM devices where magnetic interference can be mitigated~\cite{fujisaki2012overview}. The bitcell diagram of a STT-MRAM is shown in Figure~\ref{fig:nvm-fig-2}(a). 

MRAM typically requires high write currents, a source of supply noise. Deterministic supply noise can be exploited by attackers to launch DoS attacks, fault injection attacks, row hammer attacks, etc. Apart from that, MRAMs are highly susceptible to external magnetic fields. Such fields can cause the magnetic orientation of the MTJ layer to flip~\cite{ghosh2016security}, resulting in data corruption. Adversaries can take advantage of this vulnerability to execute DoS attacks. MRAM is also affected by high temperatures that can reduce data retention, and DoS attacks can be launched leveraging reduced data retention~\cite{khan2021comprehensive}.

\subsection{Resistive Random Access Memory} RRAM or ReRAM is a device with a simple metal–insulator–metal structure, where the insulator is typically an oxide of elements like Hafnium, Tantalum, or Titanium. Other materials, such as chalcogenides and 2D materials like hexagonal boron nitride, have also been used and shown in Figure~\ref{fig:nvm-fig-2}(c). RRAMs can have a single metal-insulator-metal layer or a multilayered structure, offering improved uniformity in device parameters. These devices switch between high and low resistance states, representing 1 and 0 bits. The resistive switching is achieved through SET and RESET operations, forming or rupturing conducting paths inside the insulator~\cite{gupta2020resistive}.

RRAM exhibits a critical physical vulnerability due to its filament-based switching mechanism, which is highly sensitive to current and voltage variations. The presence of parasitic capacitance ($C_p$) in a 1T1R structure can cause overshoot currents, leading to uncontrolled filament growth and reduced resistance in the low resistance state (LRS). This results in higher reset currents and prolonged switching times, causing reliability issues. Attackers could deliberately increase $C_p$ to cause instability in RRAM’s operation. Such manipulation can degrade performance, disrupt expected read/write behaviors, and potentially deplete the lifetime of RRAM~\cite{schultz2017understanding}.

\begin{figure*}[t]
    \centering
    \begin{subfigure}[b]{0.20\textwidth}
        \centering
        \includegraphics[width=\textwidth]{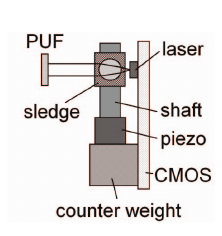}
        \caption{PCM-based PUF}
        \label{fig:puf1}
    \end{subfigure}
    \hfill
    \begin{subfigure}[b]{0.38\textwidth}
        \centering
        \includegraphics[width=\textwidth]{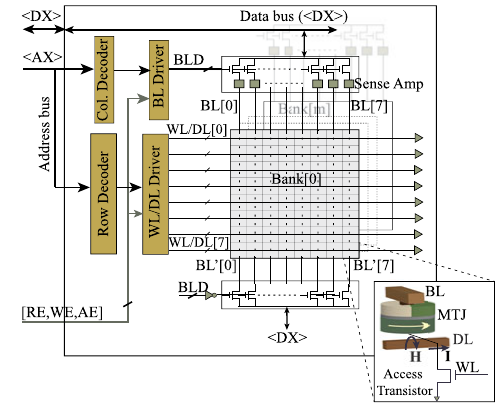}
        \caption{MRAM-based PUF}
        \label{fig:puf2}
    \end{subfigure}
    \hfill
    \begin{subfigure}[b]{0.32\textwidth}
        \centering
        \includegraphics[width=\textwidth]{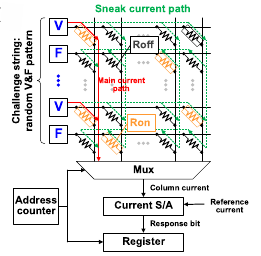}
        \caption{RRAM-based PUF}
        \label{fig:puf3}
    \end{subfigure}
    \caption{(a) Shematic side-view of a PCM-based PUF~\cite{kursawe2009reconfigurable}, (b) System level architecture of an MRAM-based PUF~\cite{das2014novel} and (c) Architecture of an RRAM-based PUF~\cite{liu2018x}.}
    \label{fig:PUFs}
    \vspace{-10px}
\end{figure*}

\subsection{Ferroelectric Random Access Memory} FeRAM, sometimes referred to as FRAM, is a type of NVM that consists of a capacitor and transistor structure (see Figure~\ref{fig:nvm-fig-2}(b)). FeRAM provides not just non-volatility but also offers fast memory access similar to DRAM~\cite{asari1999feram}. One of FeRAM's key features is its extremely low power consumption, which is unmatched by other NVM technologies like Flash. This low power requirement allows FeRAM to operate at voltages less than 2V, a significant advantage over Flash, which requires over 20V for write or erase operations. The most used ferroelectric material for FeRAMs is lead zirconate titanate (PZT)~\cite{meena2014overview}. FeRAM also offers fast writing speeds and a high number of rewrites, making it suitable for high-density and in-memory applications. These memories come in various cell types, such as capacitor, transistor, and chain cell, with the transistor type being better for high-density uses. However, this type of FeRAM has issues with data retention, not lasting 10 years in practical applications~\cite{fujisaki2012overview}.

FeRAM has several physical vulnerabilities that make it susceptible to attacks. One major issue is its asymmetric read current and high write current, which can be exploited in power analysis attacks like differential power analysis (DPA) and correlation power analysis (CPA) to extract sensitive data. Since the write current varies depending on the data being written, an attacker can analyze the patterns in power consumption to infer information. Additionally, FeRAM is vulnerable to external electric and thermal fields, which can disrupt polarization, cause data corruption, or even affect data retention. These weaknesses create opportunities for side-channel attacks and potential DoS attacks~\cite{khan2021comprehensive}.

\section{Embedded NVMs for Security}\label{sec:NVM-security} 
 
eNVMs are used to build secure architectures that are resistant to tampering and provide durability for a broad spectrum of applications, such as cryptographic key storage and secure boot processes. Furthermore, their superior scalability, reconfigurability, support for low-cost local computing, and rich source of entropy make them great candidates for security primitives like physically unclonable functions and true random number generators~\cite{Mahmoodi2019experimental}. In security applications, STT-MRAM and RRAM-based eNVMs are widely utilized. The inherent randomness in RRAM-based security systems is excellent for applications such as PUFs and TRNGs.

\vspace{5px}
\subsection{Physically Unclonable Functions}
PUFs harness residual manufacturing process variations to generate unique and unclonable device signatures~\cite{lim2005extracting}. By generating on-demand keys, PUFs eliminate the need to store keys in eNVMs during deployment, enhancing device resistance against physical attacks. These individualized keys allow for unique device identification and authentication. Memory-based PUFs, among various architectures, generate unclonable signatures without requiring hardware modifications~\cite{sutar2018memory, suh2007physical}. Considerable research has been conducted on RRAM PUFs~\cite{chen2014utilizing, mazady2015memristor}, MRAM PUFs~\cite{vatajelu2015stt,das2015mram}, PCM PUFs~\cite{zhang2013pckgen, noor2020phase}, and Flash memory PUFs~\cite{wang2012flash}. Figure~\ref{fig:PUFs} depicts representative hardware architectures used in different eNVM-based PUFs.

\subsubsection{Flash-based PUFs}PUFs leveraging flash memory emerged with Prabhu et al.~\cite{prabhu2011extracting}, who explored memory variations to generate responses but faced low throughput and security issues. Wang et al.~\cite{wang2012flash} improved efficiency using intra-page Pearson coefficients, cutting response time to 20 kb/s. A major breakthrough came when Wu et al.~\cite{wu2018puf} introduced a programming burst method that enhanced uniqueness, randomness, and resilience to environmental variations, making flash PUFs more viable for real-world applications. Mahmoodi et al.~\cite{mahmoodi2019chipsecure} further advanced the field with ChipSecure to expand the challenge-response pairs (CRP) space to resist machine learning attacks while maintaining energy efficiency. More recently, Sakib et al~\cite{sakib2020aging} refined the approach, leveraging program disturbance behavior for an aging-resistant and lightweight design suited for embedded systems.

\subsubsection{PCM-based PUFs} Kursawe et al.~\cite{kursawe2009reconfigurable} introduced reconfigurable PUFs, where memory states are weakly programmed and erased to enable dynamic response behavior. This work laid the foundation for later designs, such as multi-bit PCM-based PUFs. Figure~\ref{fig:puf1} shows a schematic side view of their PCM design, where controlled laser pulses induce weak programming to enable reconfigurable PUF behavior. Noor and Silva~\cite{noor2020phase} later identified PCM as a strong candidate for PUF applications, citing its analog resistance states, gradual programming, and suitability for reconfigurable architectures. Building on this, Zhang et al.~\cite{zhang2013pckgen} proposed PCKGen, a PCM-based reconfigurable PUF that used an imprecisely controlled current-pulse regulator to refresh cryptographic keys by injecting controlled variability during programming. To improve resistance against physical attacks, Zhang et al.~\cite{zhang2014leakage} introduced MemPUF, which performs periodic self-updates to prevent CRP reuse. While this improves unpredictability over time, secure verifier-prover communication remains an open problem.

\subsubsection{MRAM-based PUFs} Regarding the origin of MRAM-based PUFs, Marukame et al.~\cite{marukame2014extracting} proposed a method to create a PUF using MTJs in STT-MRAM. They leveraged the natural variability in MTJ switching voltages to generate a unique signature. They induced probabilistic switching by applying a controlled voltage, categorized the resistance states, and refined the selection process to extract a reliable PUF signature~\cite{gao2017emerging}. This demonstrated the feasibility of MTJ-based PUFs, though further work was needed to improve extraction reliability and consistency. Geometry-based STT-MRAM PUFs~\cite{das2014novel, das2015mram} follow a two-step process: cells are first placed in an unstable polarization state and then allowed to settle into stable configurations. Due to geometric variations in MTJ dimensions, each array produces a unique, repeatable response that can be read out as a memory PUF. Figure~\ref{fig:puf2} presents the system-level architecture of a geometry-based MRAM PUF as proposed by Das et al.~\cite{das2014novel}, in which MTJ cells are destabilized and then released to settle into unique ground states determined by intrinsic geometric variations. Other STT-MRAM PUFs rely on comparing cell resistances in the anti-parallel state~\cite{zhang2014highly, zhang2014feasibility, vatajelu2015stt, shamsi2016security}. These designs exploit TMR variation across cells to generate entropy, allowing lightweight response extraction without complex training or initialization.

\subsubsection{RRAM-based PUFs} The majority of PUF demonstrations involve the comparison of resistances among selected cells~\cite{chen2015exploiting}. This method exploits inherent process variability in resistive memory arrays, where each cell exhibits unique resistance characteristics, enabling the extraction of distinct CRPs. Early RRAM-based PUF designs~\cite{koeberl2013memristor} relied on process variations using a weak-write method to generate unique responses. These designs amplified device-level randomness by partially programming cells, producing repeatable yet device-specific outputs. While effective, they offered limited stability and lacked reconfigurability. A more robust embedded PUF~\cite{gao2015memristive} followed, enhancing flexibility without significant hardware changes. This design improved readout mechanisms and integrated more stable architectures, addressing issues related to entropy quality and environmental sensitivity. Later, Rose et al.~\cite{rose2015performance} leveraged write time variations and sneak-path currents to improve entropy extraction. By treating sneak-path interference as a usable entropy source, they demonstrated increased unpredictability and modeling resistance in crossbar-based arrays. Building on the idea of leveraging sneak-path interference, Liu et al.~\cite{liu2018x} proposed the X-point PUF, a strong RRAM-based architecture that exploits controlled sneak-path currents in cross-point arrays to expand the CRP space. Figure~\ref{fig:puf3} illustrates the architecture of an RRAM-based X-point PUF, where a random challenge pattern activates selected rows. The resulting main currents and the sneak path currents are measured through a sense amplifier to generate binary responses.

A breakthrough came with reconfigurable PUFs~\cite{lin2021highly}, where resistance fluctuations were used to dynamically refresh the challenge-response space. By resetting the memory array and reintroducing randomness post-fabrication, these designs allowed repeated regeneration of high-entropy CRPs. This method significantly reduced the bit error rate while maintaining flexibility by stochastically redistributing resistances ~\cite{zahoor2024overview}.

Zhang et al.~\cite{zhang2014feasibility} assessed the feasibility and quality of eNVM PUFs based on STT-MRAM, PCM, and RRAM. The study demonstrated that, compared to traditional memory PUFs, eNVM-based PUFs offer higher density, enabling more efficient chip area utilization for an equivalent number of bits. However, the reliability of certain eNVMs, such as RRAM-based PUFs, could potentially be influenced by reading instability and retention loss in RRAMs. Retention loss in RRAMs could additionally impact the stability of PUF-generated IDs~\cite{Chen2016true}.

\begin{figure*}[t]
    \centering
    \begin{subfigure}[b]{0.35\textwidth}
        \centering
        \includegraphics[width=\textwidth]{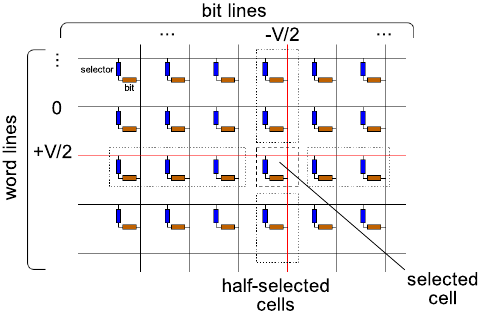}
        \caption{PCM-based TRNG}
        \label{fig:trng1}
    \end{subfigure}
    \hfill
    \begin{subfigure}[b]{0.30\textwidth}
        \centering
        \includegraphics[width=\textwidth]{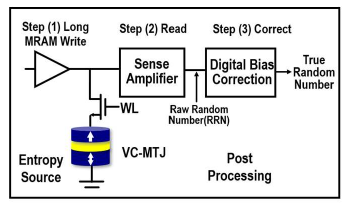}
        \caption{MRAM-based TRNG}
        \label{fig:trng2}
    \end{subfigure}
    \hfill
    \begin{subfigure}[b]{0.32\textwidth}
        \centering
        \includegraphics[width=\textwidth]{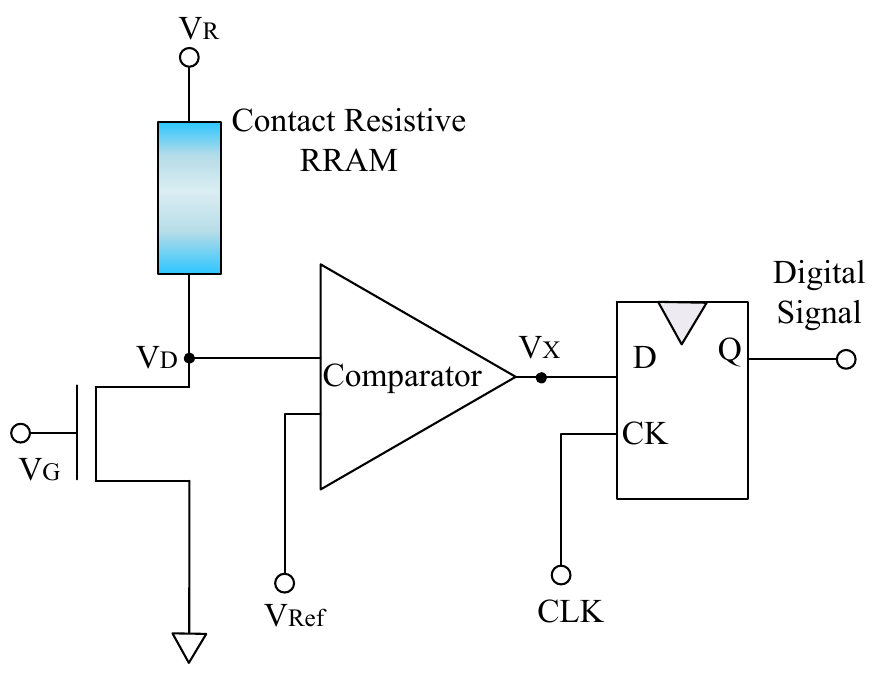}
        \caption{RRAM-based TRNG}
        \label{fig:trng3}
    \end{subfigure}
    \caption{(a) Schematic diagram of a PCM-based TRNG~\cite{piccinini2017self}, (b) block diagram of an MRAM-based TRNG~\cite{yang2021calibration} and (c) architecture of an RRAM-based TRNG~\cite{huang2012contact}.}
    \label{fig:TRNGs}
    \vspace{-10px}
\end{figure*}

\vspace{5px}
\subsection{True Random Number Generators} 
TRNGs have become integral in secure data handling systems and information security. They are crucial in generating parameters for public key cryptosystems (e.g., ECC, RSA), session keys, and many other applications. TRNGs, in contrast to pseudo-random number generators (PRNGs), derive random numbers from unpredictable physical processes, ensuring superior statistical characteristics. While PRNGs are deterministically repeatable and commonly used in simulation and testing, TRNGs offer heightened unpredictability, making them particularly suitable for applications in highly secure systems~\cite{yu2019survey}.

Extensive research on TRNGs has been conducted across various domains of non-volatile memories - spintronic devices~\cite{fukushima2014spin, liu2018spin, qu2017true}, FeRAMS~\cite{rashid2020true}, etc. The probabilistic switching nature of STT-MRAM and RRAM allows controlled programming by adjusting the pulse duration or amplitude. Experimental demonstrations have shown that, at a $50\%$ switching probability, the device has an equal chance of ending up in either the `0' or `1' state~\cite{fukushima2014spin}. This behavior can be utilized to develop TRNGs. The strong random telegraph noise (RTN) signal in RRAM can be used for random number generation in a simple circuit~\cite{huang2012contact, Chen2016true}. Examples of several TRNG architectures using different eNVM technologies are shown in Figure~\ref{fig:TRNGs}.

\subsubsection{Flash-based TRNGs} Flash memory has also been explored as a promising entropy source for true random number generation. TRNGs utilizing flash harness the intrinsic noise and variability of floating-gate memory cells to produce unpredictable bitstreams without additional hardware. Wang et al.~\cite{wang2012flash} first demonstrated that RTN in partially programmed flash cells could yield high-entropy random bits. Building on this, Ray and Milenković~\cite{ray2018true} used program-disturb stress and repeated reads to identify marginal cells prone to random flipping from RTN and read noise, enhancing randomness and leveraging aging effects. Based on partial programming and disturbance characteristics, these techniques form the foundation of flash-based TRNG designs.

\subsubsection{PCM-based TRNGs} Piccinini et al.~\cite{piccinini2017self} demonstrated the promising use of amorphous PCM arrays for implementing a TRNG in their research. Their proposed design applies a calibrated voltage pulse to a fully reset PCM array, inducing random switching through intrinsic threshold variability; this mechanism is depicted in Figure~\ref{fig:trng1}.

\subsubsection{MRAM-based TRNGs} Due to their robustness and ability to generate high-quality random numbers, MRAM-based TRNGs have garnered considerable attention. The development of MRAM-based TRNGs emerged from exploring spintronic devices as a source of randomness, such as thermal noise and dynamic variations. These devices aim to produce random numbers with high entropy and no correlation. Early works in this field~\cite{fukushima2014spin, oosawa2015design} focused on manipulating the amplitude of programming pulses to generate randomness. 
During the same period, the current-driven stochastic programming method introduced in~\cite{fong2014generating} offered a robust solution using a complementary polarizer spin dice to generate random numbers. This approach provided a more stable mechanism for TRNGs. In~\cite{vatajelu2016security}, Vatajelu et al. proposed a novel approach combining Physical PUFs and TRNGs, using MRAM by manipulating read currents for PUFs and adjusting pulse width and amplitude for TRNGs~\cite{khan2021morphable}. Yang et al.~\cite{yang2021calibration} proposed a calibration-free in-memory TRNG leveraging voltage-controlled MRAM, where randomness arises from metastable switching behavior under long write pulses, as shown in Figure~\ref{fig:trng2}.

\subsubsection{RRAM-based TRNGs} Much research has been dedicated to TRNGs based on resistive memories~\cite{huang2012contact, wei2016true, govindaraj2018csro, jiang2017novel}. TRNGs utilizing RRAMs exhibit a high entropy source, making them relatively robust and suitable for integration in high-density scenarios. Early RRAM-based TRNG efforts, such as the study by Huang et al.~\cite{huang2012contact}, exploited natural RTN in resistive devices to generate true randomness using minimal circuit complexity, illustrated in Figure~\ref{fig:trng3}. However, initial approaches to RRAM-based TRNG endeavors faced several limitations. The study by Wei et al.~\cite{wei2016true} required complex correction circuits and suffered from inconsistent noise behavior across cells, while the work in~\cite{jiang2017novel} utilizing write time variation of diffusive RRAM faced issues in speed and endurance. Lin et al.~\cite{lin2019high} later developed a high-speed and high-reliability RRAM TRNG using intrinsic analog switching characteristics. Their work enabled high throughput and robustness with minimal circuit overhead. Nevertheless, practical applications of resistive RAMs are still hindered by throughput limitations~\cite{rajendran2021application}.

\subsubsection{FeRAM-based TRNGs} TRNGs built on FeRAMs utilize the intrinsic variability of ferroelectric switching to generate high-entropy random bits efficiently. In~\cite{rashid2020true}, Rashid et al. presented a method using latency variations during write operations in commercial FeRAM chips that enable randomness extraction without external entropy sources or complex post-processing.

While aging effects in eNVMs do not compromise the randomness of TRNGs, they may lead to device degradation over time due to continuous cycling. Similarly, aging influences switching-time variability in resistive eNVM devices. It alters the threshold voltage distribution in NOR flash, which could impact device performance or the consistency of TRNG output across the lifespan~\cite{chakraborty2020true}.

\begin{table}[t]
  \caption{Summary of Security Solutions Based on eNVMs}
  \centering
  \begin{tabular}{lccc}
    \toprule
    \textbf{eNVM Technology} & \textbf{PUF} & \textbf{TRNG} & \textbf{Logic Locking} \\
    \midrule
    PCM & \cite{noor2020phase, kursawe2009reconfigurable, zhang2013pckgen} & \cite{piccinini2017self} & -- \\
    RRAM & \cite{chen2014utilizing, chen2015exploiting, rose2015performance, liu2018x} & \cite{huang2012contact, wei2016true, govindaraj2018csro, rajendran2021application} & -- \\
    MRAM & \cite{marukame2014extracting, vatajelu2015stt, das2015mram, zhang2014highly} & \cite{fukushima2014spin, fong2014generating, vatajelu2016security} & \cite{divyanshu2022logic} \\
    FLASH & \cite{prabhu2011extracting, wang2012flash, mahmoodi2019chipsecure, sakib2020aging} & \cite{wang2012flash, ray2018true} & -- \\
    FeRAM & \cite{kim2021physical} & \cite{rashid2020true} & -- \\
    \bottomrule
  \end{tabular}
  \label{tab:Security}
\end{table}

\subsection{Obfuscation and Locking}
\begin{figure}[ht]
\centering
\includegraphics[width=0.7\linewidth]{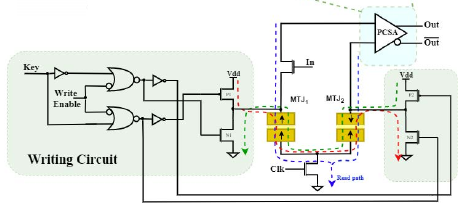} \vspace{-10px}
\caption{A logic locking block using MTJ~\cite{divyanshu2022logic}.\label{fig:logic-locking}}
\vspace{-5px}
\end{figure}

The hardware security community has actively addressed the persistent threat of IP piracy stemming from the horizontal integration of semiconductor design, manufacturing, and testing. With the growing complexity of chip design and manufacturing processes, many design houses find it practically infeasible to produce chips independently. This vulnerability in the semiconductor supply chain opens the door for untrusted entities to exploit and pirate design details, leading to irreparable damage. In response to this challenge, logic locking techniques~\cite{guin2016fortis} have been proposed as a countermeasure against IP piracy, involving the obfuscation of circuit designs through the use of secret keys. This research area remains relatively unexplored within the community, particularly in terms of integrating eNVM-based designs. The research work by Divyanshu et al.~\cite{divyanshu2022logic} explores various emerging structures based on 2T/3T MTJ for potential applications in logic locking. Figure~\ref{fig:logic-locking} illustrates their logic locking design based on a 2T STT-MTJ-based key gate. It employs complementary MTJs, a key-controlled write circuit, and a precharge sense amplifier (PCSA) for differential output evaluation. Logic inputs are applied via a CMOS block, while the key sets MTJ states that modulate resistance during evaluation by the PCSA. The design enables logic locking by enforcing key-dependent behavior under a hybrid CMOS-spintronics framework.  Additionally, magnetic skyrmion-based locking solutions were proposed by Guin et al.~\cite{zhang2022camskygate}.

Table~\ref{tab:Security} provides a summary of existing studies that utilize eNVM technologies for implementing hardware security primitives such as PUFs, TRNGs, and logic locking. It reflects the growing recognition of eNVMs as valuable building blocks for secure system design while also indicating that some technologies remain underexplored in certain application areas. As shown, RRAM and MRAM have been the most extensively studied across both PUF and TRNG implementations, highlighting their strong suitability for entropy generation and process variability-based security. MRAM is also the only technology in the table with a recorded implementation of logic locking, suggesting its potential for broader architectural integration beyond randomness-based primitives. Flash memory, while traditionally not regarded as a candidate for emerging secure designs, has demonstrated applicability in both PUF and TRNG constructs, indicating renewed research interest in repurposing legacy memory platforms for lightweight security. In contrast, PCM and FeRAM have seen more limited use. PCM has been employed in both PUFs and TRNGs, albeit with fewer studies. FeRAM appears only once in the primitives study, with no logic locking implementations to date. 

\section{Security Risks Posed by Embedded NVMs}\label{sec:NVM-attacks}

The integration of emerging eNVMs in contemporary computing systems raises significant concerns about the potential leakage of sensitive information to adversaries. Typically, confidential data like secret keys, login credentials, and credit card information undergo encryption and are stored in hard drives, such as magnetic disks or flash storage. Subsequently, this encrypted data is decrypted on-the-fly and loaded into volatile memories, such as SRAM-based caches, in close proximity to the processor. Previously, precautions were not necessary as SRAMs and DRAMs lose their content after powering down. However, implementing encryption at the cache level becomes exceedingly challenging. If cache memories become non-volatile, there is a risk of adversaries gaining access to all sensitive information in its raw form. Consequently, addressing data safety concerns in higher memory stack levels while sustaining optimal performance poses a significant challenge.

\vspace{5px}
\subsection{Side-Channel Attacks (SCA)} 
SCA poses a serious security threat to cryptographic chips used in secure systems. Unlike attacks that target the algorithm itself, SCAs focus on exploiting vulnerabilities in the physical implementation of cryptographic algorithms. eNVMs exhibit asymmetric and high read/write currents, where the currents for writing and reading data `1' and data `0' differ, making them prone to SCAs~\cite{khan2021side}. Various research works have shown that eNVM technologies such as MRAM, FeRAM, PCM, and RRAM can be susceptible to SCAs.

The work by Khan et al.~\cite{khan2017side} presented an experimental evaluation of SCAs on commercial MRAM chips. After taking the power traces of the chip, it was found that the average read current directly correlates with the Hamming Weight of the data being read, thereby confirming the presence of exploitable leakage. DPA on the read operation enabled successful key extraction with only 15 traces. This low trace requirement was attributed to reduced algorithmic noise due to byte-wise data access compared to full-word access in simulations. The CPA attack model for MRAM write~\cite{chakraborty2017correlation} also demonstrated key recovery under real system conditions, exposing the vulnerability of MRAM to SCAs. The authors showed the vulnerability of MTJ-based implementations of cryptosystems to differential side-channel attacks, in which the adversary leverages multiple traces to extract the secret key. These studies underscore the importance of accounting for magnetic switching behavior and TMR variability during both read and write phases of MRAM operation.

RRAM exhibits asymmetric read/write current behavior much like STT-MRAM. The asymmetric currents have been exploited in power analysis attacks in works like~\cite{khan2021comprehensive}. Khan et al. performed DPA targeting write operations in RRAM and retrieved the first AES key byte in approximately 900 traces, while read-based attacks required 200 traces. These attacks relied on modeling leakage using Hamming Distance and Hamming Weight, respectively. The vulnerability of IMC architectures implemented using RRAM to SCA was showcased in the study conducted by~\cite{ensan2021scare}, where power leakage during matrix-vector operations was exploited to reveal internal computations. This underscores the challenges of secure data handling in analog-mode processing using resistive memory. 

The authors of~\cite{xu2014seasoning} demonstrated SCA attacks exploiting the seasoning effect in PCM, which is the change in behavior of PCM cells as a function of operational cycles. This aging-related drift can alter the current signatures in a way that leaks information about internal states. More recently, Khan et al.~\cite{khan2021comprehensive} performed an in-depth experimental study of power side-channel vulnerabilities in PCM. Their work showed that an AES secret key could be extracted from a PCM-backed cache by collecting only a few hundred power traces during read/write operations, confirming the feasibility of real-world attacks on phase-change-based secure memory.

FeRAM is also susceptible to SCAs. Enan et al.~\cite{enan2019investigation} studied the effect of SCAs on FeRAMs with noise using signal processing techniques. Their analysis revealed that read and write operations in FeRAM produce distinguishable side-channel signatures, especially under noisy conditions, indicating exploitable leakage. 

\vspace{5px}
\subsection{Probing Attacks} 
A probing attack is an invasive technique that directly probes a signal wire to extract information from a chip using micro- or nanoprobes. During a probing attack, an adversary accesses the internal wires and connections of a targeted device to extract sensitive information. Various emerging physical probing methods can be used to gain unauthorized access or compromise the integrity of stored information in an eNVM device. STT-MRAM cells usually consist of magnetic and non-magnetic layers, placing the magnetic free layer near the middle of the device stack. This deep positioning makes it hard to directly probe the magnetic free layer non-destructively using magneto-optical current imaging (MOCI). Nonetheless, adversaries could potentially address this challenge by taking a cross-sectional image or removing stack layers until they expose the data storage layer. FeRAM may face security risks from scanning microwave impedance microscopy (sMIM) and scanning capacitance microscopy (SCM), which can detect changes in capacitance and resistance, respectively. Another potential attack method on FeRAM is electron beam-induced resistance change (EBIRCH), where changes in resistance can be measured using tools like electron beam-induced current (EBIC) or electron beam-absorbed current (EBAC) and EBIRCH. PCM could be at risk from conductive atomic force microscopy (CAFM) because it can detect the current state of the material, which varies between amorphous and crystalline states. To execute such an attack, one might need to remove layers until the active layer is exposed. RRAM employing $HfO_2$ is not susceptible to MOCI due to the absence of ferromagnetic properties. However, if NiO material with ferromagnetic properties is used, RRAM could be vulnerable to MOCI~\cite{biswas2022emerging}.

\vspace{5px}
\subsection{Fault injection (FI) Attacks} The supply noise in eNVMs, caused by high and asymmetric write currents, can be exploited for fault injection attacks. The attacker can create deterministic supply noise by writing a specific data pattern. This noise can then propagate to the memory space of the victim-user, leading to read/write operation failures. 

In~\cite{khan2018fault}, Khan et al. conducted a fault injection experiment on RRAM-based last-level cache (LLC). The high write current of RRAM can lead to supply noise, such as voltage droop and ground bounce. Their study showed that supply noise induced by high write current can be transmitted to the neighboring banks and affect parallel read/write operations. By manipulating the read/write data patterns, the attacker can influence the magnitude of the supply noise and thus execute a fault injection attack. Shortly after, Petryk et al.~\cite{petryk2020evaluation} performed the first experimental laser fault injections on actual HfO$_2$-based RRAM cells. They successfully flipped memory bits using carefully tuned optical pulses. Most recently, Kumar et al.~\cite{kumar2023fault} examined oxide RRAM under various injection methods (laser, electromagnetic pulses, and read-disturb stress). According to their results, cells in the high-resistance `OFF' state are far more vulnerable to transient faults than low-resistance `ON’ cells, which showed considerable immunity.

Fault injection attacks on flash memory often target the instruction fetch or erase operations. Skorobogatov~\cite{skorobogatov2010flash} used infrared lasers to induce bit flips in embedded flash, effectively bypassing memory protection. Colombier et al.~\cite{colombier2018laser} showed that a single laser pulse could corrupt instructions during fetch without changing stored flash contents. Viera et al.~\cite{viera2021permanent} extended this with repeated pulses to cause permanent faults in flash cells. Schink et al.~\cite{schink2024flash} established that timed glitches can suppress flash erase during boot, allowing attackers to extract protected data in the process. These attacks show that both transient and permanent faults in flash can be induced using laser or voltage glitching.

MRAM has been shown to be vulnerable to both external and internal fault injection techniques. Khan and Ghosh~\cite{khan2018fault} introduced an internal fault model where high write currents in STT-MRAM create localized voltage droops, leading to bit errors without the need for external injection. Later, Chakraborty et al.~\cite{chakraborty2022toggle} demonstrated that strong external magnetic fields can flip bits in commercial toggle MRAM chips. Yazigy et al.~\cite{yazigy2023laser} showed that infrared laser pulses can disrupt read and write operations in STT-MRAM by locally heating the memory cell. In 2024, Ahmed et al.~\cite{ahmed2024sttmram} reported that even moderate magnetic fields can corrupt data in 40nm STT-MRAM.

\vspace{5px}
\subsection{Row-hammer (RH) Attacks} The Row-hammer attack exploits electromagnetic interference to intentionally flip specific bits in DRAM memories by repetitively accessing particular rows. These intentional bit-flips violate an important rule in secure computing called memory isolation. This rule ensures a strict separation of application memory to prevent unauthorized changes in its internal state. Few studies have investigated the impact of Rowhammer on eNVMs such as STT-MRAM. The reduced thermal barrier in STT-MRAM could result in retention failures and make the bits sensitive to stray magnetic fields and thermal noise. Researchers in~\cite{khan2018analysis} investigated the effects of Row-hammer attacks on STT-MRAM using high write current. The effects of this attack on STT-MRAM are not as severe as DRAM, but it can create different types of failures and affect more bit cells. At the same time,  Row-hammer attacks can result in retention problems and read disturb issues if read operations are conducted while cells experience disturbed current due to ground bounce. Such attacks also introduce the possibility of read/write failures. 

In 2022, Staudigl et al.~\cite{staudigl2022neurohammer} demonstrated a rowhammer-style attack on RRAM (memristor crossbars). Their experimental results showed that repeatedly writing to selected cells on shared word or bit lines could flip bits in unselected, neighboring cells. This was attributed to cumulative stress in half-selected RRAM cells due to voltage and thermal coupling. The paper confirmed that the effect was reproducible and effective in both simulation and hardware prototypes, marking a concrete realization of rowhammer in the RRAM domain.

Rowhammer-like effects have been demonstrated in MLC NAND flash, where repeated file system-level operations induce bit flips in adjacent cells through program interference. IBM researchers~\cite{kurmus2017there} showed that such access patterns can exploit threshold voltage shifts to corrupt neighboring data, revealing a new attack surface in flash-based SSDs.

\subsection{Information Leakage (IL) Attacks} 
Information leakage attacks on eNVMs exploit their physical and electrical characteristics to infer sensitive data. Due to factors like high write currents and asymmetric access behavior, memory operations can produce detectable side effects, such as supply noise, which can unintentionally reveal information about sensitive data. An adversary may extract partial knowledge of the memory contents without direct access by monitoring these effects.

In~\cite{khan2018information}, Khan et al. described an information leakage attack on embedded RRAM, where an adversary exploits supply noise generated by a victim’s write operation to infer sensitive data. High and asymmetric write currents in eNVMs cause voltage droop and ground bounce, propagating through shared power networks. By performing rapid reads on nearby memory regions, the adversary can detect read failures correlated with the victim’s write activity and estimate the Hamming Weight of the victim’s data. The authors also stated that while their experimental modeling is based on RRAM, the attack methodology applies to other eNVMs, including STT-MRAM. In their 2019 study, Kommareddy et al.~\cite{kommareddy2019crossbar} showed that memristor-based crossbar arrays exhibit content-dependent write latency due to sneak path currents, introducing a new class of information leakage channels. In their WRITE+TIME attack model, a malicious process manipulates the resistive state of its memory cells to modulate the write latency and enable covert communication. Additionally, Khan and Ghosh~\cite{khan2021comprehensive} identified information leakage vulnerabilities in STT-MRAM and MRAM arising from data-dependent write and read currents.

\subsection{Denial of Service Attacks} A DoS attack on memory disrupts legitimate access by overwhelming or destabilizing memory resources without altering stored data. In eNVMs, adversaries can issue repeated high-current writes to induce supply voltage fluctuations, triggering read/write failures in adjacent cells and effectively denying service to users.

The study by Khan et al.~\cite{khan2021comprehensive} presented a simulation-based investigation demonstrating the feasibility of DoS attacks on RRAM by exploiting supply noise-induced failures. Specifically, the authors model a scenario in which an adversary and a victim share an LLC so that the adversary can induce deterministic supply voltage droop and ground bounce through carefully crafted high-current write operations. These noise artifacts propagate through the power grid and impact the victim's memory accesses. The results show that when the combined voltage loss at the victim's bitcell exceeds 120 mV, complete write failures occur, manifesting as a DoS attack. The study further characterizes polarity-dependent write failures, observing that a supply noise range of 50-120 mV can selectively prevent LRS to HRS switching to enable 0 to 1 fault injection. While read-induced noise is also evaluated, the results indicate that inducing read errors for data `0' requires significantly higher noise. The research work by Arafin et al.~\cite{arafin2020Security} highlighted that in processing-in-memory (PIM) systems built on eNVMs such as RRAM and STT-MRAM, maintaining atomicity of in-memory operations exposes new attack surfaces. In particular, adversaries may launch DoS attacks by corrupting PIM directory entries. Such manipulations can delay or block legitimate read/write operations, causing significant service disruption.

\subsection{Thermal Attacks} The temperature sensitivity of eNVMs can be exploited to launch thermal attacks. Almost all types of eNVMs are susceptible to such attacks~\cite{khan2021comprehensive}. An attacker can accelerate charge leakage or trigger resistance drift in memory cells by increasing the ambient temperature or applying localized heating. These thermal effects can shift cell states from their intended values, causing bit-flips and leading to read or write failures in memories.

\begin{table}[t]
  \caption{Summary of Risks Associated with eNVMs}
  \centering
  \begin{tabular}{lccccccc}
    \toprule
    \textbf{eNVM Technology} & \textbf{SCA} & \textbf{Probing} & \textbf{FI} & \textbf{RH} & \textbf{IL} & \textbf{DoS} & \textbf{Thermal} \\
    \midrule
    PCM & \cite{xu2014seasoning} & \cite{kannan2014security,biswas2022emerging} & -- & -- & -- & -- & \cite{boybat2021temperature} \\
    RRAM & \cite{ensan2021scare} & \cite{biswas2022emerging} & \cite{khan2018fault, kumar2023fault, petryk2020evaluation} & \cite{staudigl2022neurohammer} & \cite{khan2018information} & \cite{khan2021comprehensive, arafin2020Security} & \cite{staudigl2024s} \\
    MRAM & \cite{chakraborty2017correlation, khan2017side} & \cite{biswas2022emerging} & \cite{khan2018fault, yazigy2023laser} & \cite{khan2018analysis, agarwal2018rowhammer} & \cite{khan2021comprehensive} & \cite{arafin2020Security} & \cite{jang2016performance} \\
    FLASH & -- & -- & \cite{colombier2018laser, viera2021permanent} & \cite{kurmus2017there} & -- & -- & -- \\
    FeRAM & \cite{enan2019investigation} & \cite{biswas2022emerging} & -- & -- & -- & -- & -- \\
    \bottomrule
  \end{tabular}
  \label{tab:Attacks}
\end{table}

STT-MRAM is sensitive to thermal manipulation due to the temperature dependence of magnetization dynamics in its MTJ structure. Jang et al.~\cite{jang2016performance} demonstrated that heating the chip reduces the retention time and sense margin by degrading parameters like saturation magnetization and polarization. Elevated temperature increases the likelihood of read disturb and accelerates spontaneous bit flipping. Their simulations showed that heating and cooling can lead to performance and security failures by subtly shifting write, read, and retention characteristics.
RRAMs are vulnerable to thermal manipulation due to temperature-dependent ion migration within their switching filaments. Staudigl et al.~\cite{staudigl2022neurohammer, staudigl2024s} demonstrated this with NeuroHammer, a thermal crosstalk-based attack that induces bit-flips by heating adjacent cells through repeated switching. This localized heating accelerates the switching dynamics from high to low-resistance states.

PCM-based analog in-memory computing is susceptible to temperature fluctuations, which can induce resistance drift and unintended bit flips. Studies like~\cite{boybat2021temperature} have shown that such thermal variations degrade computational accuracy by shifting resistance values away from their intended states. These findings reveal a potential attack surface where thermal manipulation could compromise data integrity in PCM systems.

Table \ref{tab:Attacks} shows the summary of security attacks and vulnerabilities posed by eNVMs. It highlights the diverse range of threats targeting different memory technologies. It is important to note that this table reflects only the presence of published studies. The absence of a specific attack category for a given memory type does not indicate immunity, but suggests that the vulnerability may remain underexplored or insufficiently studied in current literature. Among the technologies listed, MRAM and RRAM are linked to the broadest range of documented threats. This diversity likely reflects the extensive security research focused on these memories and their intrinsic device properties, such as stochastic switching in MRAM and analog resistance variation in RRAM, making them attractive targets for attack modeling. PCM is also associated with several known vulnerabilities, particularly fault injection, probing, and thermal attacks. These threats can be largely attributed to its sensitivity to temperature and phase-change thresholds. Flash appears in fault injection and row-hammer studies, consistent with its high-density charge storage architecture. FeRAM, in contrast, is minimally represented across the attack categories. This limited coverage may not imply robustness but rather highlights a gap in literature that warrants further investigation.

\section{Research Trends}\label{sec:nvm-trends}

\begin{figure}[htbp]
\centering
\includegraphics[width=0.85\linewidth]{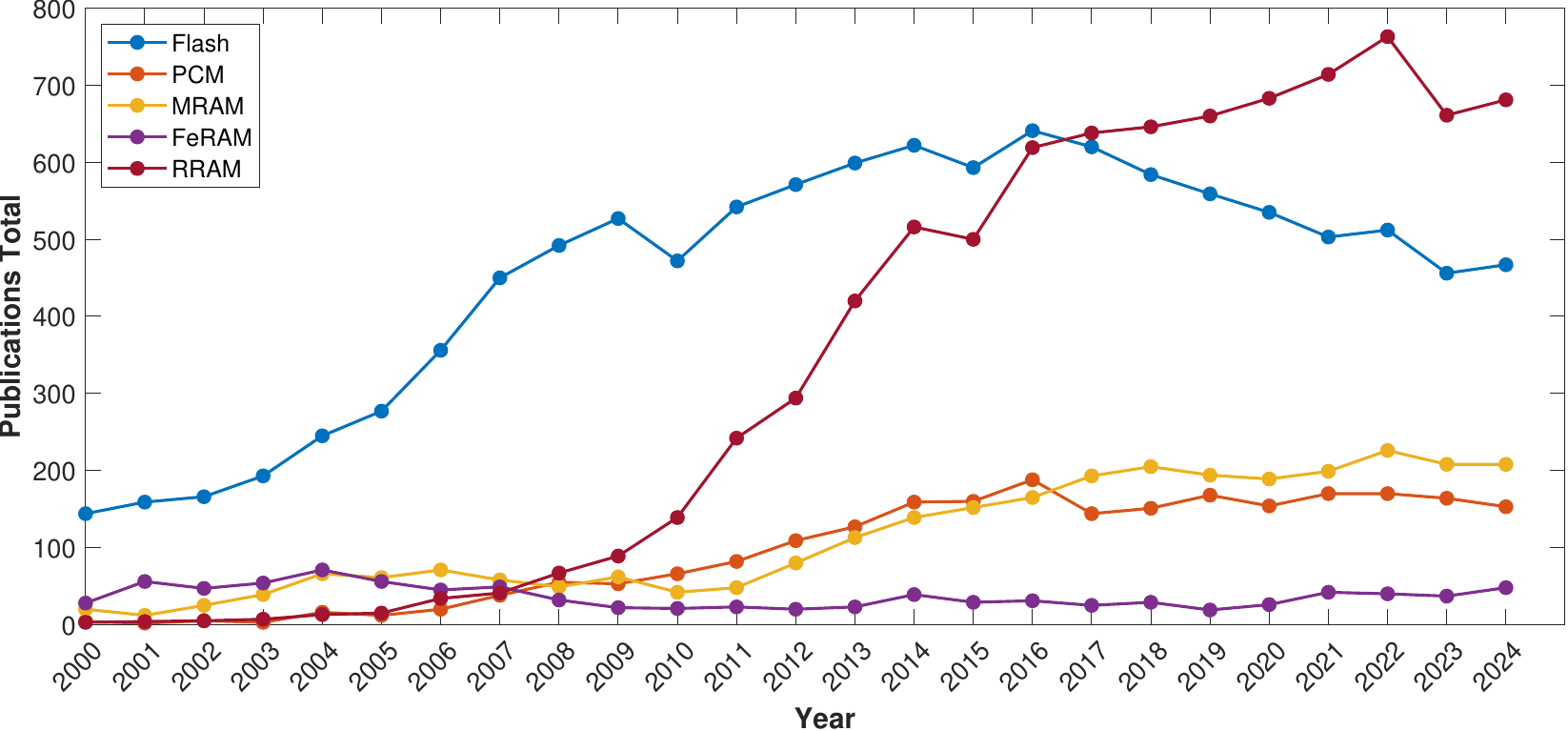} 
\caption{Annual publication trends for non-volatile memory from 2000 to 2024.}  \label{fig:nvm-trends}
\vspace{0px}
\end{figure}

The growing interest in NVMs is evident in the rising volume of research within the field, reflecting a broader push toward finding improved memory solutions for future technologies. Figure~\ref{fig:nvm-trends} illustrates the overall increase of research publications focusing on the five major NVM technologies since the early 2000s. The data presented in this figure have been collected from the Web of Science database. Flash dominated the early 2000s, reaching a peak around 2014 with more than 600 publications. A gradual decline followed, likely driven by scaling limitations and increasing maturity in conventional applications. In contrast, RRAM and MRAM have experienced significant growth in research activity. RRAM, in particular, showed a rapid increase beginning around 2010 and surpassed Flash in annual publication count by 2020, highlighting its emergence as a scalable and adaptable memory candidate. MRAM has also demonstrated steady growth, particularly after 2013, corresponding to progress in STT and SOT-based implementations. These shifts suggest a clear transition in research focus toward more advanced and application-specific NVM technologies.

\begin{figure}[t]
\centering
\includegraphics[width=0.5\linewidth]{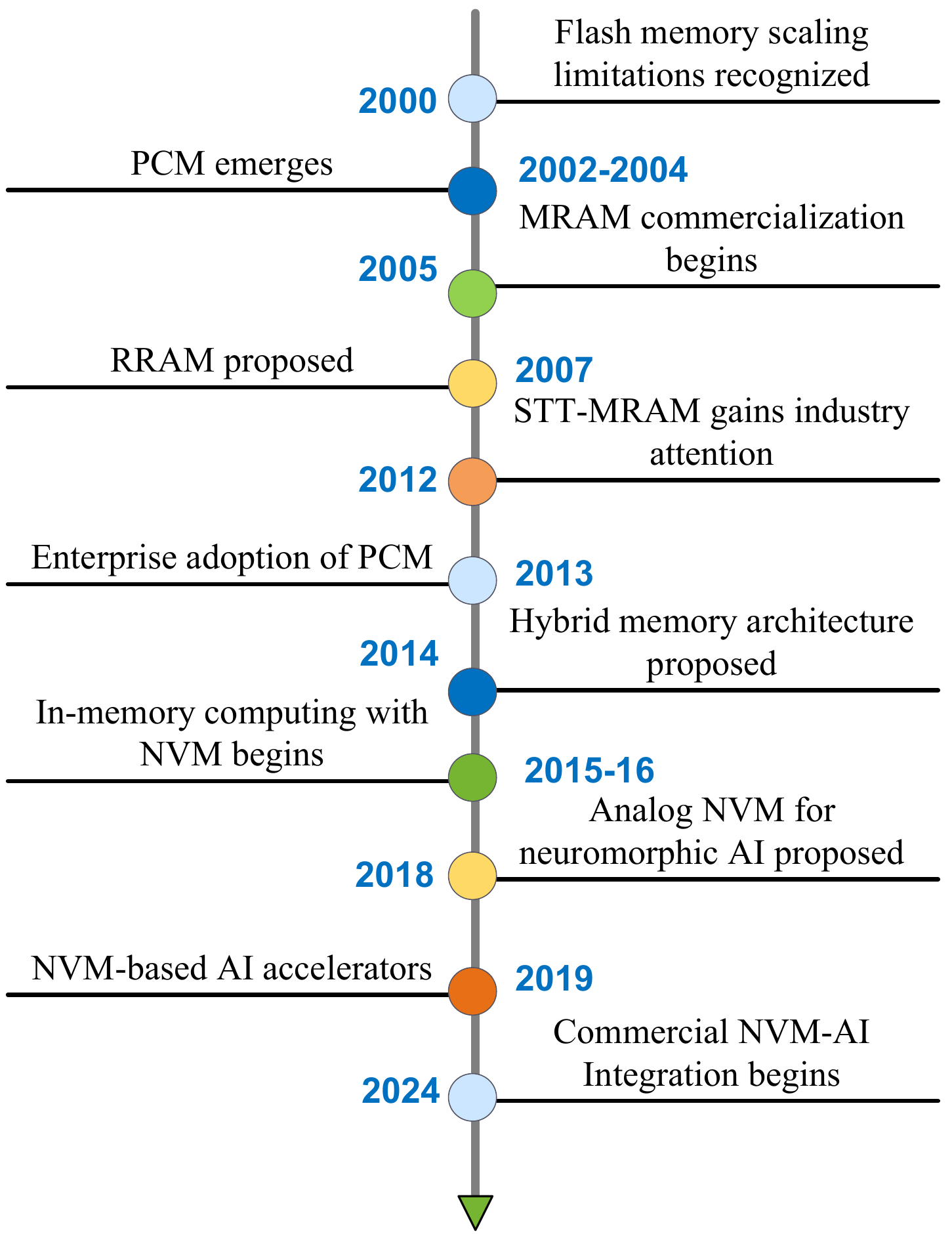} \vspace{-10px}
\caption{ Key milestones in the evolution of NVM technologies.\label{fig:timeline}}
\vspace{-10px}
\end{figure}

This rising research activity corresponds closely with key technological milestones that have shaped the evolution of NVMs. Figure~\ref{fig:timeline} presents a timeline of major innovations and their integration into computing systems over the past two decades. In the early 2000s, the semiconductor community began recognizing the inherent scaling limitations of Flash, primarily due to physical constraints such as charge leakage and cell-to-cell interference. This realization sparked interest in alternative memory paradigms. Between 2002 and 2004, PCM emerged as a viable candidate offering multilevel cell capability. PCM was later adopted in enterprise-grade storage solutions due to its reliability and density advantages. In 2005, MRAM entered the commercial arena, combining non-volatility with fast read/write performance, making it suitable for embedded applications. By 2007, RRAM was proposed, offering a simple two-terminal configuration with analog switching properties. These characteristics render RRAM particularly attractive for both scalable memory and neuromorphic applications.

The timeline continued with the initiation of 3D NAND development in 2011, marking a significant architectural shift to overcome planar NAND density limitations. In 2012, STT-MRAM began attracting substantial industrial attention due to its high CMOS compatibility. Subsequently, 2013 witnessed growing enterprise interest in PCM. As these technologies matured, hybrid memory architectures were explored in 2014 to combine the speed of DRAM with the persistence of NVM~\cite{salkhordeh2016operating}. This convergence laid the groundwork for in-memory computing, which gained momentum between 2015 and 2016, allowing data processing within memory arrays and reducing latency and energy costs. In 2018, analog variants of NVM, such as RRAM and PCM, found utility in neuromorphic computing platforms, serving as adaptive synaptic weights in brain-inspired systems. This progression paved the way for NVM-based AI accelerators in 2019, optimized for parallel data processing and machine learning tasks. By 2024, the commercial integration of NVM-AI platforms had been achieved, marking a notable milestone in the convergence of NVM technologies and intelligent computing architectures.

\section{Future Directions}\label{sec:NVM-future}

As eNVMs gain traction in modern computing, their integration into the next-generation memory architecture for intelligent computing offers several advantages. In particular, AI accelerators for in-memory computing with RRAM and PCM support parallel and low-power matrix operations, helping to reduce data movement and improve energy efficiency in edge systems~\cite{wan2022compute}. Architectures like SOT-MRAM that feature separate read and write paths help lower dynamic energy during frequent accesses.  In addition to these architectural benefits, eNVM technologies are known for their high density and compatibility with advanced CMOS nodes, making them well-suited for compact and efficient SoC integration~\cite{pentecost2021nvmexplorer}. However, while eNVMs scale well at the device level, large-scale array deployment in applications like deep learning introduces challenges. These include interconnect complexity, leakage currents, and variability in resistance states, especially at deep sub-micron nodes. Such limitations can compromise the reliability of memory-intensive systems and necessitate careful cross-layer design considerations~\cite{yu2016binary}. Looking ahead, the integration of eNVMs into AI accelerators is expected to play a central role in the evolution of edge and neuromorphic computing platforms. As AI workloads grow in complexity and scale, memory technologies such as RRAM, PCM, and STT-MRAM offer promising capabilities for supporting localized and parallel computation while minimizing data movement bottlenecks. In the near term, advances including 3D-stacked NVM arrays, hybrid CMOS-eNVM compute fabrics, and multilevel resistance encoding are projected to significantly enhance computational density and energy efficiency. Beyond these architectural innovations, future directions may involve co-designing AI models and eNVM-based hardware to leverage intrinsic device characteristics for intelligent and context-aware processing. Collectively, these developments are expected to reshape the architecture of memory systems for next-generation intelligent computing.

Furthermore, other limitations can impact the broader applicability of eNVMs in computing and security domains. Notably, endurance remains a concern in write-heavy workloads such as those found in deep neural network training, where frequent updates can degrade devices like STT-MRAM, PCM, and RRAM~\cite{shirinzadeh2017endurance}. Compared to SRAM, eNVMs tend to have higher write latency and energy consumption. This creates a trade-off where the benefits of non-volatility come at the expense of reduced performance in real-time applications. Future work may focus on improving endurance and write efficiency, so that eNVMs can more effectively support applications with high performance and frequent update demands.

In parallel with architectural advancements, the growing adoption of eNVMs raises concerns regarding their security resilience. There is an increasing need to assess their vulnerability to various threats at both the device and system levels, such as fault injection from peripheral interfaces. System-level features like wear-leveling could also be exploited for attacks. The security community should consider new attack vectors beyond denial of service and fault injection. Areas such as information leakage in eNVMs require more attention due to the many ways they can be exploited for data leakage. A multilayered approach is needed to address these challenges to enhance eNVM security against physical attacks. One can integrate nanopyramid structures and protective shields at the device level to protect against optical attacks. Material-based approaches like antiferromagnetic materials and superconductors can also be explored. Carbon nanotube resistance sensors can also be integrated for real-time tamper detection~\cite{biswas2022emerging}.

Beyond physical tamper protection, ensuring the authenticity of the eNVM hardware itself is critical. Counterfeit detection techniques that identify recycled or cloned memory chips should be developed to secure the hardware supply chain. These methods must be lightweight and architecture-aware to minimize performance and area overhead. Alongside such preventative strategies, one of the most pressing challenges in mitigating eNVM attacks is developing efficient detection methods. Future research approaches can explore new device engineering techniques that reduce vulnerabilities without adding significant overhead. As post-quantum cryptography continues to gain momentum, future research could examine the potential of eNVMs as secure storage platforms for quantum-resistant algorithms. This includes evaluating whether their endurance and retention capabilities can support the frequent key updates and computational demands associated with post-quantum cryptographic schemes.

\section{Conclusion}\label{sec:conclusion}
As eNVMs become essential components in modern computing systems, their security implications require dedicated and ongoing attention. This paper has provided a comprehensive analysis of both the capabilities and vulnerabilities associated with eNVM security. In addition to reviewing their architectural foundations, we have identified key design factors contributing to their susceptibility to security threats. Our study also presents a detailed discussion of current research trends in eNVM-based security solutions. It is evident from the literature that eNVMs serve as promising candidates for building security primitives such as PUFs and TRNGs. However, the same features that enable these applications also expose eNVMs to a wide range of physical and logical attacks. We have discussed several attack vectors highlighting the spectrum of security threats targeting eNVMs across different technologies, including information leakage, denial-of-service, and thermal attacks. Furthermore, this work includes an analysis of publication trends and technological advancements in the eNVM domain over the past two decades. We hope this study serves as a comprehensive reference for researchers seeking to understand both the security benefits and the challenges associated with eNVM integration in secure system architectures.

\section*{Acknowledgment} This work was supported by the National Science Foundation under Grant Number CNS-2312139.

\bibliographystyle{unsrt}
\bibliography{NonVolatileMemory}

\begin{thebibliography}{100}

\bibitem{khan2021comprehensive}
Mohammad Nasim~Imtiaz Khan and Swaroop Ghosh.
\newblock {Comprehensive Study of Security and Privacy of Emerging Non-Volatile Memories}.
\newblock {\em Journal of low power electronics and applications}, 11(4):36, 2021.

\bibitem{tisha2024exploring}
Zakia~Tamanna Tisha, Jeremy Muldavin, and Ujjwal Guin.
\newblock {Exploring Security Solutions and Vulnerabilities for Embedded Non-Volatile Memories}.
\newblock In {\em IEEE Computer Society Annual Symposium on VLSI}, pages 361--366, 2024.

\bibitem{derhacobian2010power}
Narbeh Derhacobian, Shane~C Hollmer, Nad Gilbert, and Michael~N Kozicki.
\newblock {Power and Energy Perspectives of Nonvolatile Memory Technologies}.
\newblock {\em Proceedings of the IEEE}, (2):283--298, 2010.

\bibitem{gerardin2010present}
Simone Gerardin and Alessandro Paccagnella.
\newblock {Present and Future Non-Volatile Memories for Space}.
\newblock {\em IEEE Transactions on nuclear science}, 57(6):3016--3039, 2010.

\bibitem{vatajelu2014nonvolatile}
Elena~Ioana Vatajelu, Hassen Aziza, and Cristian Zambelli.
\newblock {Nonvolatile memories: Present and future challenges}.
\newblock In {\em Int. Design and Test Symposium}, pages 61--66, 2014.

\bibitem{aswathy2021future}
N~Aswathy and NM~Sivamangai.
\newblock {Future Nonvolatile Memory Technologies: Challenges and Applications}.
\newblock In {\em International Conference on Advances in Computing, Communication, Embedded \& Secure Systems}, pages 308--312, 2021.

\bibitem{bez2003introduction}
Roberto Bez, Emilio Camerlenghi, Alberto Modelli, and Angelo Visconti.
\newblock {Introduction to flash memory}.
\newblock {\em Proceedings of the IEEE}, (4):489--502, 2003.

\bibitem{meena2014overview}
Jagan~Singh Meena, Simon~Min Sze, Umesh Chand, and Tseung-Yuen Tseng.
\newblock {Overview of emerging nonvolatile memory technologies}.
\newblock {\em Nanoscale research letters}, pages 1--33, 2014.

\bibitem{fujisaki2012overview}
Yoshihisa Fujisaki.
\newblock {Overview of emerging semiconductor non-volatile memories}.
\newblock {\em IEICE Electronics Express}, (10):908--925, 2012.

\bibitem{wang2019threat}
Gang Wang.
\newblock {Threat Models and Security of Phase-Change Memory}.
\newblock In {\em 2019 IEEE International Conference on Consumer Electronics}, pages 1--6, 2019.

\bibitem{ghosh2016security}
Swaroop Ghosh, Mohammad Nasim~Imtiaz Khan, Asmit De, and Jae-Won Jang.
\newblock {Security and privacy threats to on-chip non-volatile memories and countermeasures}.
\newblock In {\em 2016 IEEE/ACM International Conference on Computer-Aided Design (ICCAD)}, pages 1--6, 2016.

\bibitem{khan2021study}
Mohammad Nasim~Imtiaz Khan and Swaroop Ghosh.
\newblock {Comprehensive Study of Security and Privacy of Emerging Non-Volatile Memories}.
\newblock {\em Journal of low power electronics and applications}, (4):36, 2021.

\bibitem{gupta2020resistive}
Varshita Gupta, Shagun Kapur, Sneh Saurabh, and Anuj Grover.
\newblock {Resistive Random Access Memory: A Review of Device Challenges}.
\newblock {\em IETE Technical Review}, (4):377--390, 2020.

\bibitem{schultz2017understanding}
Thomas Schultz and Rashmi Jha.
\newblock {Understanding vulnerabilities in ReRAM devices for trust in semiconductor designs}.
\newblock In {\em 2017 IEEE National Aerospace and Electronics Conference (NAECON)}, pages 338--342, 2017.

\bibitem{kursawe2009reconfigurable}
Klaus Kursawe, Ahmad-Reza Sadeghi, Dries Schellekens, Boris Skoric, and Pim Tuyls.
\newblock {Reconfigurable Physical Unclonable Functions - Enabling technology for tamper-resistant storage}.
\newblock In {\em 2009 IEEE International Workshop on Hardware-Oriented Security and Trust}, pages 22--29, 2009.

\bibitem{das2014novel}
Jayita Das, Kevin Scott, Drew Burgett, Srinath Rajaram, and Sanjukta Bhanja.
\newblock {A novel geometry based MRAM PUF}.
\newblock In {\em Int. Conference on Nanotechnology}, pages 859--863, 2014.

\bibitem{liu2018x}
Rui Liu, Pai-Yu Chen, Xiaochen Peng, and Shimeng Yu.
\newblock {X-Point PUF: Exploiting Sneak Paths for a Strong Physical Unclonable Function Design}.
\newblock {\em IEEE Transactions on Circuits and Systems I: Regular Papers}, 65(10):3459--3468, 2018.

\bibitem{asari1999feram}
K.~Asari, Y.~Mitsuyama, T.~Onoye, I.~Shirakawa, H.~Hirano, T.~Honda, T.~Otsuki, T.~Baba, and T.~Meng.
\newblock {FeRAM circuit technology for system on a chip}.
\newblock In {\em Proceedings of the First NASA/DoD Workshop on Evolvable Hardware}, pages 193--197, 1999.

\bibitem{Mahmoodi2019experimental}
M.~R. Mahmoodi, D.~B. Strukov, and O.~Kavehei.
\newblock {Experimental Demonstrations of Security Primitives With Nonvolatile Memories}.
\newblock {\em Transactions on Electron Devices}, (12):5050--5059, 2019.

\bibitem{lim2005extracting}
Daihyun Lim, Jae~W Lee, Blaise Gassend, G~Edward Suh, Marten Van~Dijk, and Srinivas Devadas.
\newblock {Extracting secret keys from integrated circuits}.
\newblock {\em IEEE Transactions on Very Large Scale Integration Systems}, (10):1200--1205, 2005.

\bibitem{sutar2018memory}
Soubhagya Sutar, Arnab Raha, and Vijay Raghunathan.
\newblock {Memory-Based Combination PUFs for Device Authentication in Embedded Systems}.
\newblock {\em IEEE Transactions on Multi-Scale Computing Systems}, (4):793--810, 2018.

\bibitem{suh2007physical}
G~Edward Suh and Srinivas Devadas.
\newblock {Physical Unclonable Functions for Device Authentication and Secret Key Generation}.
\newblock In {\em Proceedings of the 44th annual design automation conference}, pages 9--14, 2007.

\bibitem{chen2014utilizing}
An~Chen.
\newblock {Utilizing the Variability of Resistive Random Access Memory to Implement Reconfigurable Physical Unclonable Functions}.
\newblock {\em IEEE Electron Device Letters}, (2):138--140, 2014.

\bibitem{mazady2015memristor}
Anas Mazady, Md~Tauhidur Rahman, Domenic Forte, and Mehdi Anwar.
\newblock {Memristor PUF—A Security Primitive: Theory and Experiment}.
\newblock {\em IEEE Journal on Emerging and Selected Topics in Circuits and Systems}, (2):222--229, 2015.

\bibitem{vatajelu2015stt}
Elena~Ioana Vatajelu, Giorgio Di~Natale, Marco Indaco, and Paolo Prinetto.
\newblock {STT MRAM-based PUFs}.
\newblock In {\em Design, Automation \& Test in Eu. Conference \& Exhibition}, pages 872--875, 2015.

\bibitem{das2015mram}
Jayita Das, Kevin Scott, Srinath Rajaram, Drew Burgett, and Sanjukta Bhanja.
\newblock {MRAM PUF: A Novel Geometry Based Magnetic PUF With Integrated CMOS}.
\newblock {\em IEEE Transactions on Nanotechnology}, (3):436--443, 2015.

\bibitem{zhang2013pckgen}
Le~Zhang, Zhi~Hui Kong, and Chip-Hong Chang.
\newblock {PCKGen: A Phase Change Memory based cryptographic key generator}.
\newblock In {\em International Symposium on Circuits and Systems}, pages 1444--1447, 2013.

\bibitem{noor2020phase}
Nafisa Noor and Helena Silva.
\newblock {Phase Change Memory for Physical Unclonable Functions}.
\newblock {\em Applications of Emerging Memory Technology: Beyond Storage}, pages 59--91, 2020.

\bibitem{wang2012flash}
Yinglei Wang, Wing-kei Yu, Shuo Wu, Greg Malysa, G.~Edward Suh, and Edwin~C. Kan.
\newblock {Flash Memory for Ubiquitous Hardware Security Functions: True Random Number Generation and Device Fingerprints}.
\newblock In {\em Symposium on Security \& Privacy}, pages 33--47, 2012.

\bibitem{prabhu2011extracting}
Pravin Prabhu, Ameen Akel, Laura~M Grupp, Wing-Kei~S Yu, G~Edward Suh, Edwin Kan, and Steven Swanson.
\newblock {Extracting Device Fingerprints from Flash Memory by Exploiting Physical Variations}.
\newblock In {\em Trust and Trustworthy Computing: International Conference}, pages 188--201. Springer, 2011.

\bibitem{wu2018puf}
Meng-Yi Wu, Tsao-Hsin Yang, Lun-Chun Chen, Chi-Chang Lin, Hao-Chun Hu, Fang-Ying Su, Chih-Min Wang, James Po-Hao Huang, Hsin-Ming Chen, Chris Chun-Hung Lu, et~al.
\newblock {A PUF scheme using competing oxide rupture with bit error rate approaching zero}.
\newblock In {\em 2018 IEEE International Solid-State Circuits Conference-(ISSCC)}, pages 130--132, 2018.

\bibitem{mahmoodi2019chipsecure}
Mohammad Mahmoodi, Hussein Nili, Shabnam Larimian, Xinjie Guo, and Dmitri Strukov.
\newblock {ChipSecure: A Reconfigurable Analog eFlash-Based PUF with Machine Learning Attack Resiliency in 55nm CMOS}.
\newblock In {\em Proceedings of the 56th Annual Design Automation Conference 2019}, pages 1--6, 2019.

\bibitem{sakib2020aging}
Sadman Sakib, Aleksandar Milenkovi{\'c}, Md~Tauhidur Rahman, and Biswajit Ray.
\newblock {An Aging-Resistant NAND Flash Memory Physical Unclonable Function}.
\newblock {\em IEEE Transactions on Electron Devices}, 67(3):937--943, 2020.

\bibitem{zhang2014leakage}
Le~Zhang, Chip-Hong Chang, Alessandro Cabrini, Guido Torelli, and Zhi~Hui Kong.
\newblock {Leakage-resilient memory-based physical unclonable function using phase change material}.
\newblock In {\em 2014 International Carnahan Conference on Security Technology (ICCST)}, pages 1--6, 2014.

\bibitem{marukame2014extracting}
Takao Marukame, Tetsufumi Tanamoto, and Yuichiro Mitani.
\newblock {Extracting Physically Unclonable Function From Spin Transfer Switching Characteristics in Magnetic Tunnel Junctions}.
\newblock {\em IEEE Transactions on Magnetics}, 50(11):1--4, 2014.

\bibitem{gao2017emerging}
Yansong Gao, Damith~C Ranasinghe, Said~F Al-Sarawi, Omid Kavehei, and Derek Abbott.
\newblock {Emerging Physical Unclonable Functions With Nanotechnology}.
\newblock {\em IEEE access}, 4:61--80, 2017.

\bibitem{zhang2014highly}
Le~Zhang, Xuanyao Fong, Chip-Hong Chang, Zhi~Hui Kong, and Kaushik Roy.
\newblock {Highly reliable memory-based Physical Unclonable Function using Spin-Transfer Torque MRAM}.
\newblock In {\em Int. symposium on circuits \& systems}, pages 2169--2172, 2014.

\bibitem{zhang2014feasibility}
Le~Zhang, Xuanyao Fong, Chip-Hong Chang, Zhi~Hui Kong, and Kaushik Roy.
\newblock {Feasibility study of emerging non-volatile memory based physical unclonable functions}.
\newblock In {\em Int. Memory Workshop}, pages 1--4, 2014.

\bibitem{shamsi2016security}
Kaveh Shamsi and Yier Jin.
\newblock {Security of emerging non-volatile memories: Attacks and defenses}.
\newblock In {\em VLSI Test Symposium}, pages 1--4, 2016.

\bibitem{chen2015exploiting}
Pai-Yu Chen, Runchen Fang, Rui Liu, Chaitali Chakrabarti, Yu~Cao, and Shimeng Yu.
\newblock {Exploiting resistive cross-point array for compact design of physical unclonable function}.
\newblock In {\em Int. Symposium on Hardware Oriented Security and Trust}, pages 26--31, 2015.

\bibitem{koeberl2013memristor}
Patrick Koeberl, {\"U}nal Kocaba{\c{s}}, and Ahmad-Reza Sadeghi.
\newblock {Memristor PUFs: A new generation of memory-based Physically Unclonable Functions}.
\newblock In {\em Design, Automation \& Test in Eu. Conference}, pages 428--431, 2013.

\bibitem{gao2015memristive}
Yansong Gao, Damith~C Ranasinghe, Said~F Al-Sarawi, Omid Kavehei, and Derek Abbott.
\newblock {Memristive crypto primitive for building highly secure physical unclonable functions}.
\newblock {\em Scientific reports}, 5(1):12785, 2015.

\bibitem{rose2015performance}
Garrett~S Rose and Chauncey~A Meade.
\newblock {Performance analysis of a memristive crossbar PUF design}.
\newblock In {\em Proceedings of the 52nd Annual Design Automation Conference}, pages 1--6, 2015.

\bibitem{lin2021highly}
Bohan Lin, Yachuan Pang, Bin Gao, Jianshi Tang, Dong Wu, Ting-Wei Chang, Wei-En Lin, Xiaoyu Sun, Shimeng Yu, Meng-Fan Chang, et~al.
\newblock {A Highly Reliable RRAM Physically Unclonable Function Utilizing Post-Process Randomness Source}.
\newblock {\em IEEE Journal of Solid-State Circuits}, pages 1641--1650, 2021.

\bibitem{zahoor2024overview}
Furqan Zahoor, Arshid Nisar, Usman~Isyaku Bature, Haider Abbas, Faisal Bashir, Anupam Chattopadhyay, Brajesh~Kumar Kaushik, Ali Alzahrani, and Fawnizu~Azmadi Hussin.
\newblock {An overview of critical applications of resistive random access memory}.
\newblock {\em Nanoscale Advances}, 2024.

\bibitem{Chen2016true}
An~Chen.
\newblock {A review of emerging non-volatile memory (NVM) technologies and applications}.
\newblock {\em Solid-State Electronics}, pages 25--38, 2016.

\bibitem{piccinini2017self}
Enrico Piccinini, Rossella Brunetti, and Massimo Rudan.
\newblock {Self-Heating Phase-Change Memory-Array Demonstrator for True Random Number Generation}.
\newblock {\em IEEE Transactions on Electron Devices}, (5):2185--2192, 2017.

\bibitem{yang2021calibration}
Jiyue Yang, Di~Wu, Albert Lee, Seyed~Armin Razavi, Puneet Gupta, Kang~L Wang, and Sudhakar Pamarti.
\newblock {A Calibration-Free In-Memory True Random Number Generator Using Voltage-Controlled MRAM}.
\newblock In {\em IEEE 51st European Solid-State Device Research Conference}, pages 115--118, 2021.

\bibitem{huang2012contact}
Chien-Yuan Huang, Wen~Chao Shen, Yuan-Heng Tseng, Ya-Chin King, and Chrong-Jung Lin.
\newblock {A Contact-Resistive Random-Access-Memory-Based True Random Number Generator}.
\newblock {\em Electron Device Letters}, pages 1108--1110, 2012.

\bibitem{yu2019survey}
Fei Yu, Lixiang Li, Qiang Tang, Shuo Cai, Yun Song, and Quan Xu.
\newblock {A Survey on True Random Number Generators Based on Chaos}.
\newblock {\em Discrete Dynamics in Nature and Society}, pages 1--10, 2019.

\bibitem{fukushima2014spin}
Akio Fukushima, Takayuki Seki, Kay Yakushiji, Hitoshi Kubota, Hiroshi Imamura, Shinji Yuasa, and Koji Ando.
\newblock {Spin dice: A scalable truly random number generator based on spintronics}.
\newblock {\em Applied Physics Express}, (8):083001, 2014.

\bibitem{liu2018spin}
Yang Liu, Zhaohao Wang, Zuwei Li, Xiaoxiao Wang, and Weisheng Zhao.
\newblock {A spin orbit torque based true random number generator with real-time optimization}.
\newblock In {\em 18th International Conference on Nanotechnology}, pages 1--4, 2018.

\bibitem{qu2017true}
Yuanzhuo Qu, Jie Han, Bruce~F Cockburn, Witold Pedrycz, Yue Zhang, and Weisheng Zhao.
\newblock {A true random number generator based on parallel STT-MTJs}.
\newblock In {\em Design, Automation \& Test in Eu. Conference \& Exhibition}, pages 606--609, 2017.

\bibitem{rashid2020true}
Md~Imtiaz Rashid, Farah Ferdaus, BMS~Bahar Talukder, Paul Henny, Aubrey~N Beal, and Md~Tauhidur Rahman.
\newblock {True Random Number Generation Using Latency Variations of FRAM}.
\newblock {\em IEEE Transactions on Very Large Scale Integration Systems}, (1):14--23, 2020.

\bibitem{ray2018true}
Biswajit Ray and Aleksandar Milenkovi{\'c}.
\newblock {True Random Number Generation Using Read Noise of Flash Memory Cells}.
\newblock {\em IEEE transactions on electron devices}, (3):963--969, 2018.

\bibitem{oosawa2015design}
Satoshi Oosawa, Takayuki Konishi, Naoya Onizawa, and Takahiro Hanyu.
\newblock {Design of an STT-MTJ based true random number generator using digitally controlled probability-locked loop}.
\newblock In {\em 2015 IEEE 13th International New Circuits and Systems Conference (NEWCAS)}, pages 1--4, 2015.

\bibitem{fong2014generating}
Xuanyao Fong, Mei-Chin Chen, and Kaushik Roy.
\newblock {Generating true random numbers using on-chip complementary polarizer spin-transfer torque magnetic tunnel junctions}.
\newblock In {\em Device Research Conference}, pages 103--104, 2014.

\bibitem{vatajelu2016security}
Elena~Ioana Vatajelu, Giorgio Di~Natale, and Paolo Prinetto.
\newblock {Security primitives (PUF and TRNG) with STT-MRAM}.
\newblock In {\em IEEE 34th VLSI Test Symposium}, pages 1--4, 2016.

\bibitem{khan2021morphable}
Mohammad Nasim~Imtiaz Khan, Chak~Yuen Cheng, Sung~Hao Lin, Abdullah Ash-Saki, and Swaroop Ghosh.
\newblock {A Morphable Physically Unclonable Function and True Random Number Generator using a Commercial Magnetic Memory}.
\newblock {\em Journal of Low Power Electronics and Applications}, (1):5, 2021.

\bibitem{wei2016true}
Z~Wei, Y~Katoh, S~Ogasahara, Y~Yoshimoto, K~Kawai, Y~Ikeda, K~Eriguchi, K~Ohmori, and S~Yoneda.
\newblock {True random number generator using current difference based on a fractional stochastic model in 40-nm embedded ReRAM}.
\newblock In {\em International Electron Devices Meeting}, pages 4--8, 2016.

\bibitem{govindaraj2018csro}
Rekha Govindaraj, Swaroop Ghosh, and Srinivas Katkoori.
\newblock {CSRO-Based Reconfigurable True Random Number Generator Using RRAM}.
\newblock {\em IEEE Transactions on VLSI Systems}, (12):2661--2670, 2018.

\bibitem{jiang2017novel}
Hao Jiang, Daniel Belkin, Sergey~E Savel’ev, Siyan Lin, Zhongrui Wang, Yunning Li, Saumil Joshi, Rivu Midya, Can Li, Mingyi Rao, et~al.
\newblock {A novel true random number generator based on a stochastic diffusive memristor}.
\newblock {\em Nature communications}, 8(1):882, 2017.

\bibitem{lin2019high}
Bohan Lin, Bin Gao, Yachuan Pang, Peng Yao, Dong Wu, Hu~He, Jianshi Tang, He~Qian, and Huaqiang Wu.
\newblock {A High-Speed and High-Reliability TRNG Based on Analog RRAM for IoT Security Application}.
\newblock In {\em IEEE International Electron Devices Meeting}, pages 14--8, 2019.

\bibitem{rajendran2021application}
Gokulnath Rajendran, Writam Banerjee, Anupam Chattopadhyay, and Mohamed M~Sabry Aly.
\newblock {Application of Resistive Random Access Memory in Hardware Security: A Review}.
\newblock {\em Advanced Electronic Materials}, (12):2100536, 2021.

\bibitem{chakraborty2020true}
Supriya Chakraborty, Abhilash Garg, and Manan Suri.
\newblock {True Random Number Generation From Commodity NVM Chips}.
\newblock {\em IEEE Transactions on Electron Devices}, (3):888--894, 2020.

\bibitem{divyanshu2022logic}
Divyanshu Divyanshu, Rajat Kumar, Danial Khan, Selma Amara, and Yehia Massoud.
\newblock {Logic Locking Using Emerging 2T/3T Magnetic Tunnel Junctions for Hardware Security}.
\newblock {\em IEEE Access}, pages 102386--102395, 2022.

\bibitem{kim2021physical}
Sihyun Kim, Kitae Lee, Min-Hye Oh, Jong-Ho Lee, Byung-Gook Park, and Daewoong Kwon.
\newblock {Physical Unclonable Functions Using Ferroelectric Tunnel Junctions}.
\newblock {\em IEEE Electron Device Letters}, (6):816--819, 2021.

\bibitem{guin2016fortis}
Ujjwal Guin, Qihang Shi, Domenic Forte, and Mark~M Tehranipoor.
\newblock {FORTIS: A Comprehensive Solution for Establishing Forward Trust for Protecting IPs and ICs}.
\newblock {\em ACM Transactions on Design Automation of Electronic Systems (TODAES)}, 21(4):1--20, 2016.

\bibitem{zhang2022camskygate}
Yuqiao Zhang, Chunli Tang, Peng Li, and Ujjwal Guin.
\newblock Camskygate: camouflaged skyrmion gates for protecting ics.
\newblock In {\em Proceedings of the 59th ACM/IEEE Design Automation Conference}, pages 757--762, 2022.

\bibitem{khan2021side}
Mohammad Nasim~Imtiaz Khan, Shivam Bhasin, Bo~Liu, Alex Yuan, Anupam Chattopadhyay, and Swaroop Ghosh.
\newblock {Comprehensive Study of Side-Channel Attack on Emerging Non-Volatile Memories}.
\newblock {\em Journal of Low Power Electronics and Applications}, (4):38, 2021.

\bibitem{khan2017side}
Mohammad Nasim~Imtiaz Khan, Shivam Bhasin, Alex Yuan, Anupam Chattopadhyay, and Swaroop Ghosh.
\newblock {Side-Channel Attack on STTRAM Based Cache for Cryptographic Application}.
\newblock In {\em Int. Conference on Computer Design}, pages 33--40, 2017.

\bibitem{chakraborty2017correlation}
Abhishek Chakraborty, Ankit Mondal, and Ankur Srivastava.
\newblock {Correlation power analysis attack against STT-MRAM based cyptosystems}.
\newblock {\em Cryptology ePrint Archive}, 2017.

\bibitem{ensan2021scare}
Sina~Sayyah Ensan, Karthikeyan Nagarajan, Mohammad Nasim~Imtiaz Khan, and Swaroop Ghosh.
\newblock {SCARE: Side Channel Attack on In-Memory Computing for Reverse Engineering}.
\newblock {\em Transactions on VLSI Systems}, (12):2040--2051, 2021.

\bibitem{xu2014seasoning}
Lei Xu, Weidong Shi, and Nicholas Desalvo.
\newblock {Seasoning effect based side channel attacks to AES implementation with Phase Change Memory}.
\newblock In {\em Proceedings of the Third Workshop on Hardware and Architectural Support for Security and Privacy}, pages 1--8, 2014.

\bibitem{enan2019investigation}
Abyad Enan and Mohammed Imamul~Hassan Bhuiyan.
\newblock {Investigation of Side Channel Leakage of FeRAM Using Discrete Wavelet Transform}.
\newblock In {\em International Conference on Telecommunications and Photonics}, pages 1--4, 2019.

\bibitem{biswas2022emerging}
Liton~Kumar Biswas, M~Shafkat, M~Khan, Leonidas Lavdas, and Navid Asadizanjani.
\newblock {Emerging Nonvolatile Memories—An Assessment of Vulnerability to Probing Attacks}.
\newblock In {\em ISTFA}, pages 217--224, 2022.

\bibitem{khan2018fault}
Mohammad Nasim~Imtiaz Khan and Swaroop Ghosh.
\newblock {Fault injection attacks on emerging non-volatile memory and countermeasures}.
\newblock In {\em Proceedings of the 7th International Workshop on Hardware and Architectural Support for Security and Privacy}, pages 1--8, 2018.

\bibitem{petryk2020evaluation}
Dmytro Petryk, Zoya Dyka, Eduardo Perez, Mamathamba~Kalishettyhalli Mahadevaiaha, Ievgen Kabin, Christian Wenger, and Peter Langend{\"o}rfer.
\newblock {Evaluation of the Sensitivity of RRAM Cells to Optical Fault Injection Attacks}.
\newblock In {\em 23rd Euromicro Conference on Digital System Design}, pages 238--245, 2020.

\bibitem{kumar2023fault}
Ankit Kumar, Robin Degraeve, Arthur Beckers, Andrea Fantini, Ingrid Verbauwhede, Dimitri Linten, and Gouri~S Kar.
\newblock {Fault Attack Investigation on TaO x Resistive-RAM for Cyber Secure Application}.
\newblock {\em IEEE Transactions on Electron Devices}, 70(8):4170--4177, 2023.

\bibitem{skorobogatov2010flash}
Sergei Skorobogatov.
\newblock {Flash Memory ‘Bumping’ Attacks}.
\newblock In {\em International Workshop on Cryptographic Hardware and Embedded Systems}, pages 158--172. Springer, 2010.

\bibitem{colombier2018laser}
Baptiste Colombier, Alain Menu, Jean-Max Dutertre, Paul-Alain Mo{\"e}llic, Jean-Baptiste Rigaud, and Jean-Luc Danger.
\newblock {Laser-induced Single-bit Faults in Flash Memory: Instructions Corruption on a 32-bit Microcontroller}.
\newblock {\em IACR Cryptology ePrint Archive}, 2018:1042, 2018.

\bibitem{viera2021permanent}
Romain A.~C. Viera, Jean-Max Dutertre, and Paul-Alain Mo{\"e}llic.
\newblock {Permanent Laser Fault Injection into the Flash Memory of a Microcontroller}.
\newblock In {\em 2021 IEEE 19th International New Circuits and Systems Conference (NEWCAS)}, pages 1--4, 2021.

\bibitem{schink2024flash}
Martin Schink, Andreas Wagner, Florian Oberhansl, Simon K{\"o}ckeis, Erwin Strieder, et~al.
\newblock {Unlock the Door to my Secrets, but don't Forget to Glitch: A Comprehensive Analysis of Flash Erase Suppression Attacks}.
\newblock {\em IACR Transactions on Cryptographic Hardware and Embedded Systems}, 2024(2):88--129, 2024.

\bibitem{chakraborty2022toggle}
Subhadeep Chakraborty and Manan Suri.
\newblock {Experimental Study of Adversarial Magnetic Field Exposure Attacks on Toggle MRAM Chips}.
\newblock {\em IEEE Transactions on Electron Devices}, 69(3):1480--1485, 2022.

\bibitem{yazigy2023laser}
Nadim Yazigy, Jean Postel-Pellerin, Valerio De~Marca, Ricardo~C. Sousa, Gael Di~Pendina, and Paul Canet.
\newblock {Correlation between 1064 nm laser attack and thermal behavior in STT-MRAM}.
\newblock {\em Microelectronics Reliability}, 150:115167, 2023.

\bibitem{ahmed2024sttmram}
Md~Shoaib Ahmed, Biplab Ray, et~al.
\newblock {Assessing Magnetic Attack on Commercial 40 nm pMTJ STT-MRAM}.
\newblock In {\em IEEE International Conference on Physical Assurance and Inspection of Electronics (PAINE)}, 2024.

\bibitem{khan2018analysis}
Mohammad Nasim~Imtiaz Khan and Swaroop Ghosh.
\newblock {Analysis of Row Hammer Attack on STTRAM}.
\newblock In {\em International Conference on Computer Design}, pages 75--82, 2018.

\bibitem{staudigl2022neurohammer}
Felix Staudigl, Hazem Al~Indari, Daniel Sch{\"o}n, Dominik Sisejkovic, Farhad Merchant, Jan~Moritz Joseph, Vikas Rana, Stephan Menzel, and Rainer Leupers.
\newblock {NeuroHammer: Inducing Bit-Flips in Memristive Crossbar Memories}.
\newblock In {\em Design, Automation \& Test in Europe Conference \& Exhibition}, pages 1181--1184, 2022.

\bibitem{kurmus2017there}
Anil Kurmus, Nikolas Ioannou, Matthias Neugschwandtner, Nikolaos Papandreou, and Thomas Parnell.
\newblock {Is there a “rowhammer” for MLC NAND Flash SSDs? An analysis of filesystem attack vectors}.
\newblock In {\em USENIX Workshop on Offensive Technologies co-located with USENIX Security}, 2017.

\bibitem{khan2018information}
Mohammad Nasim~Imtiaz Khan and Swaroop Ghosh.
\newblock {Information Leakage Attacks on Emerging Non-Volatile Memory and Countermeasures}.
\newblock In {\em Proceedings of the International Symposium on Low Power Electronics and Design}, pages 1--6, 2018.

\bibitem{kommareddy2019crossbar}
Vamsee~Reddy Kommareddy, Baogang Zhang, Fan Yao, Rickard Ewetz, and Amro Awad.
\newblock {Are Crossbar Memories Secure? New Security Vulnerabilities in Crossbar Memories}.
\newblock {\em IEEE Computer Architecture Letters}, 18(2):174--177, 2019.

\bibitem{arafin2020Security}
Md~Tanvir Arafin and Zhaojun Lu.
\newblock Security challenges of processing-in-memory systems.
\newblock In {\em Proceedings of the 2020 on Great Lakes Symposium on VLSI (GLSVLSI '20)}, pages 229--234. ACM, September 2020.

\bibitem{kannan2014security}
Sachhidh Kannan, Naghmeh Karimi, Ozgur Sinanoglu, and Ramesh Karri.
\newblock {Security Vulnerabilities of Emerging Nonvolatile Main Memories and Countermeasures}.
\newblock {\em IEEE Transactions on Computer-Aided Design of Integrated Circuits and Systems}, (1):2--15, 2014.

\bibitem{boybat2021temperature}
Irem Boybat, Benedikt Kersting, S~Ghazi Sarwat, X~Timoneda, Robert~L Bruce, Matthew BrightSky, Manuel Le~Gallo, and Abu Sebastian.
\newblock {Temperature sensitivity of analog in-memory computing using phase-change memory}.
\newblock In {\em IEEE International Electron Devices Meeting}, pages 28--3, 2021.

\bibitem{staudigl2024s}
Felix Staudigl, Hazem Al~Indari, Daniel Sch{\"o}n, Hsin-Yu Chen, Dominik Sisejkovic, Jan~Moritz Joseph, Vikas Rana, Stephan Menzel, Amelie Hagelauer, and Rainer Leupers.
\newblock {It's Getting Hot in Here: Hardware Security Implications of Thermal Crosstalk on ReRAMs}.
\newblock {\em IEEE Transactions on Reliability}, 2024.

\bibitem{agarwal2018rowhammer}
S~Agarwal, H~Dixit, D~Datta, M~Tran, D~Houssameddine, D~Shum, and F~Benistant.
\newblock {Rowhammer for Spin Torque based Memory: Problem or not?}
\newblock In {\em International Magnetics Conference}, pages 1--1, 2018.

\bibitem{jang2016performance}
Jae-Won Jang and Swaroop Ghosh.
\newblock {Performance Impact of Magnetic and Thermal Attack on STTRAM and Low-Overhead Mitigation Techniques }.
\newblock In {\em Proceedings of the 2016 International Symposium on Low Power Electronics and Design}, pages 136--141, 2016.

\bibitem{salkhordeh2016operating}
Reza Salkhordeh and Hossein Asadi.
\newblock {An operating system level data migration scheme in hybrid DRAM-NVM memory architecture}.
\newblock In {\em 2016 Design, Automation \& Test in Europe Conference \& Exhibition (DATE)}, pages 936--941, 2016.

\bibitem{wan2022compute}
Weier Wan, Rajkumar Kubendran, Clemens Schaefer, Sukru~Burc Eryilmaz, Wenqiang Zhang, Dabin Wu, Stephen Deiss, Priyanka Raina, He~Qian, Bin Gao, et~al.
\newblock {A compute-in-memory chip based on resistive random-access memory}.
\newblock {\em Nature}, 608(7923):504--512, 2022.

\bibitem{pentecost2021nvmexplorer}
Lillian Pentecost, Alexander Hankin, Marco Donato, Mark Hempstead, Gu-Yeon Wei, and David Brooks.
\newblock {NVMExplorer: A Framework for Cross-Stack Comparisons of Embedded Non-Volatile Memories}.
\newblock {\em arXiv preprint arXiv:2109.01188}, 2021.

\bibitem{yu2016binary}
Shimeng Yu, Zhiwei Li, Pai-Yu Chen, Huaqiang Wu, Bin Gao, Deli Wang, Wei Wu, and He~Qian.
\newblock {Binary Neural Network with 16 Mb RRAM Macro Chip for Classification and Online Training}.
\newblock In {\em 2016 IEEE International Electron Devices Meeting (IEDM)}, pages 16--2, 2016.

\bibitem{shirinzadeh2017endurance}
Saeideh Shirinzadeh, Mathias Soeken, Pierre-Emmanuel Gaillardon, Giovanni De~Micheli, and Rolf Drechsler.
\newblock {Endurance Management for Resistive Logic-In-Memory Computing Architectures}.
\newblock In {\em Design, Automation \& Test in Europe Conference \& Exhibition (DATE), 2017}, pages 1092--1097, 2017.

\end{thebibliography}

\end{document}